\documentclass[a4paper,fleqn]{cas-dc}

\usepackage[numbers]{natbib}

\graphicspath{{}}
\usepackage{color, colortbl}
\usepackage{chemformula}
\usepackage{xcolor}
\definecolor{Gray}{gray}{0.9}

\usepackage{color, colortbl}
\usepackage{chemformula}
\usepackage{xcolor}
\usepackage{adjustbox}
\usepackage{mathrsfs}
\usepackage{booktabs}
\usepackage{amssymb}
\usepackage{hyperref}
\usepackage{pifont}
\usepackage{subcaption}

\begin{document}

\shorttitle{Local manifold learning and its link to domain-based physics knowledge}
\shortauthors{Zdyba\l{} et~al.}

\title [mode = title]{Local manifold learning and its link to domain-based physics knowledge}

\author[ULB,BURN]{Kamila Zdyba\l{}*}
\author[ULB,BURN,POLIMI]{Giuseppe D'Alessio}
\author[UE]{Antonio Attili}
\author[ULB,BURN]{Axel Coussement}
\author[UoU]{James C. Sutherland}
\author[ULB,BURN]{Alessandro Parente}

\address[ULB]{Université Libre de Bruxelles, École polytechnique de Bruxelles, Aero-Thermo-Mechanics Laboratory, Brussels, Belgium}
\address[BURN]{Université Libre de Bruxelles and Vrije Universiteit Brussel, Combustion and Robust Optimization Group (BURN), Brussels, Belgium}
\address[POLIMI]{CRECK Modeling Lab, Department of Chemistry, Materials and Chemical Engineering, Politecnico di Milano, Piazza Leonardo da Vinci 32, 20133 Milano, Italy}
\address[UE]{University of Edinburgh, School of Engineering, Institute of Multiscale Thermofluids, The King’s Buildings, Mayfield Road, Edinburgh, EH9 3FD, United Kingdom}
\address[UoU]{Department of Chemical Engineering, University of Utah, Salt Lake City, Utah, USA}

\begin{abstract}
In many reacting flow systems, the thermo-chemical state-space is known or assumed to evolve close to a low-dimensional manifold (LDM). Various approaches are available to obtain those manifolds and subsequently express the original high-dimensional space with fewer parameterizing variables. Principal component analysis (PCA) is one of the dimensionality reduction methods that can be used to obtain LDMs. PCA does not make prior assumptions about the parameterizing variables and retrieves them empirically from the training data. In this paper, we show that PCA applied in local clusters of data (local PCA) is capable of detecting the intrinsic parameterization of the thermo-chemical state-space. We first demonstrate that utilizing three common combustion models of varying complexity: the Burke-Schumann model, the chemical equilibrium model and the homogeneous reactor. Parameterization of these models is known \textit{a priori} which allows for benchmarking with the local PCA approach. We further extend the application of local PCA to a more challenging case of a turbulent non-premixed $n$-heptane/air jet flame for which the parameterization is no longer obvious. Our results suggest that meaningful parameterization can be obtained also for more complex datasets. We show that local PCA finds variables that can be linked to local stoichiometry, reaction progress and soot formation processes. \\
\\
* Corresponding author: \texttt{Kamila.Zdybal@ulb.ac.be}
\end{abstract}


\maketitle

\section{Introduction} 

Parameterization methods for modeling turbulent reacting flows rely on expressing the entire thermo-chemical state-space with fewer variables \citep{pope1992, van2000modelling, gicquel2000, van2002modelling, sutherland2007quantitative, pradeep2012}. This allows for an easier treatment of combustion process that is in general high-dimensional due to many species involved in chemical reactions. Recently, dimensionality reduction and manifold learning techniques have been used to infer low-dimensional parameterizations directly from the training data \citep{sutherland2009combustion, yang2013empirical, coussement2013}. In particular, principal component analysis (PCA) identifies the new parameters as directions of the largest variance in the ambient space. The new parameters, called principal components (PCs), are linear combinations of the preprocessed thermo-chemical state variables.

The question then arises if the parameters obtained in a data-driven way can be given any physical meaning. If PCA is able to retrieve physical information about the combustion process without any \textit{a priori} knowledge, this can yield useful implications for modeling complex systems using PCs. For instance, progress variables can be identified automatically, without the need for human insight \cite{najafiyazdi2012}. Evidence from previous research \citep{parente2011investigation, isaac2015, ranade2019apriori, malik2020combustion, d2021feature} suggests that physical meaning can still be attributed to PCs. A common way to interpret PCs is to assess coefficients of the linear combination that formed them. Since those coefficients can be negative or positive, PCA can potentially differentiate between reactants and products. Results of Parente et al. \cite{parente2011investigation} and Malik et al. \cite{malik2020combustion} suggest that PCA can identify the mixture fraction variable, $f$, as defined by Bilger et al. \cite{bilger1990}. The PC that can be given the meaning of $f$ is characterized by high coefficients of opposite signs for fuel and oxidizer components. Such PC is thus a compact representation of mixture stoichiometry, even though $f$ is not explicitly included as one of the variables in the training data.

While some insights are available from globally applied manifold learning, it is less clear what parameterizations can be derived from applying dimensionality reduction locally. For instance, if PCA is applied locally in identified zones (or clusters) of state-space, local parameterizations can be obtained, tied to specific regions of the flame. Cluster analysis for physically-relevant characteristics has been the concern of various researchers in the domain of reacting flows \citep{parente2011investigation, fooladgar2018new, dalessio2020analysis, d2020impact, li2021study}. Previous research demonstrated that local PCs can vary significantly between clusters and can be linked to physical processes associated with local regions \cite{parente2011investigation, dalessio2020analysis, li2021study}. 

The goal of this work is to interpret the low-dimensional parameterization defined by the local PCs. We adopt a bottom-up approach and test local PCA on combustion datasets of gradually increasing complexity. We first show that applying PCA in local clusters identified by the vector quantization PCA (VQPCA) algorithm \cite{kambhatla1997dimension}, we are able to retrieve parameters inherent to the construction of common combustion models: Burke-Schumann (BS), chemical equilibrium (EQ) and homogeneous reactor (HR). All three models assume that compositions lie close to a low-dimensional manifold (LDM) and the thermo-chemical state-space can be parameterized with one variable. We then test the local PCA approach on a more challenging case of a turbulent non-premixed $n$-heptane/air jet flame dataset coming from direct numerical simulation (DNS). The DNS dataset used here allows to link the local PCs to various phenomena beyond what the analysis of BS, EQ and HR datasets allows for, such as soot formation processes \citep{attili2014formation, attili2016effects, dalessio2020analysis, d2020unsupervised}. We further show how findings carry over between the simpler strained laminar flamelet (SLF) model and the DNS dataset, which includes more physical scatter than the simple combustion models.

\section{The analyzed datasets} \label{sec:data-sets} 

We first apply local PCA on three combustion models with \textit{a priori} known parameterization:
\begin{enumerate}
\item The Burke-Schumann (BS) model for methane/air with temperature and 4 species mass fractions (mass fraction of \ch{N2} is removed from PCA). 10k observations are generated for 10k equally-spaced points in the $f$ space. For $f=0$ and $f=1$, the pure stream temperature is 300K.
\item The chemical equilibrium (EQ) model for methane/air with temperature and 51 species mass fractions (mass fractions of \ch{N2} and \ch{Ar} are removed from PCA) using species from the GRI-3.0 chemical mechanism \cite{gri30}. 10k observations are generated for 10k equally-spaced points in the $f$ space. For $f=0$ and $f=1$, the pure stream temperature is 300K and pressure is 101325Pa.
\item The adiabatic, isobaric, open homogeneous reactor (HR) for hydrogen/air with temperature and 8 species mass fractions (mass fractions of \ch{N2}, \ch{Ar} and \ch{He} are removed from PCA). The initial and feed temperature is $1000$K and pressure is $101325$Pa. This dataset has 538 observations, generated on a temporal grid.
\end{enumerate}
In the supplementary material, we apply local PCA to the EQ and HR datasets with additional fuels to cover hydrogen/air, syngas/air and methane/air combustion for both datasets. Hydrogen/air and methane/air combustion is used with the same settings as mentioned for datasets 2 and 3. When syngas/air combustion is studied, we use the chemical mechanism by Hawkes et al. \cite{hawkes2007scalar} and the state-space is formed from temperature and 10 species mass fractions (mass fraction of \ch{N2} is removed from PCA). In the supplement, we show that the conclusions reported in the main text still hold for other fuels.

We then demonstrate our approach on a high-fidelity simulation dataset for which parameterization is no longer obvious:
\begin{enumerate}
\setcounter{enumi}{3}
\item The direct numerical simulation (DNS) of a 3D turbulent non-premixed $n$-heptane/air jet flame at a Reynolds number of around 15,000 and with no significant extinction. The dataset represents one 2D plane obtained from a 3D domain at $t=15$ms and is composed of temperature and 46 species mass fractions (mass fraction of \ch{N2} is removed from PCA) based on the reduced mechanism by Bisetti et al. \cite{bisetti2012on}. This data has approximately 260k observations. More information about the dataset can be found in \citep{attili2014formation}.
\end{enumerate}
We make a connection between the findings from the DNS dataset and the one-dimensional flame case corresponding to the DNS dataset:
\begin{enumerate}
\setcounter{enumi}{4}
\item The strained laminar flamelet (SLF) model for $n$-heptane/air with temperature and 46 species mass fractions (mass fraction of \ch{N2} is removed from PCA) based on the reduced mechanism by Bisetti et al. \cite{bisetti2012on}. We use two variations of the SLF dataset: one with unity Lewis number (Le) assumption (approximately 21k observations) and one with a mixture-averaged differential diffusion model (approximately 11k observations). More information about the dataset can be found in \citep{attili2016effects}. 
\end{enumerate}

Each dataset 1-5 is represented by a matrix $\mathbf{X} \in \mathbb{R}^{N \times Q}$, where $N$ is the number of observations and $Q$ is the number of state variables. This matrix defines the thermo-chemical state-space, $\mathbf{X} = \big[ T, Y_1, Y_2, \dots, Y_{n_s} \big]$, where $T \in \mathbb{R}^{N \times 1}$ is the temperature vector, $Y_i \in \mathbb{R}^{N \times 1}$ is the mass fraction of species $i$ vector and $n_s$ is the number of linearly independent species mass fractions. We refer to the $i^{\text{th}}$ observation (row) in $\mathbf{X}$ as $X_{i*}$ and to the $j^{\text{th}}$ variable (column) in $\mathbf{X}$ as $X_{*j}$. 

Before applying dimensionality reduction, the original dataset has to be preprocessed (centered and scaled). The preprocessing can be performed as $\widetilde{\mathbf{X}} = (\mathbf{X} - \mathbf{C}) \mathbf{D}^{-1}$, 
where $\mathbf{C} \in \mathbb{R}^{N \times Q}$ is the matrix of centers (typically mean values of each variable) and $\mathbf{D} \in \mathbb{R}^{Q \times Q}$ is the diagonal matrix of scales. Each element, $d_j$, on the diagonal of $\mathbf{D}$ is the scaling factor for the $j^{\text{th}}$ variable. In this work, we test two data preprocessing scenarios:
\begin{itemize}
\item Auto scaling, where $d_j = \text{std}(X_{*j})$ with the state vector defined as $\mathbf{X} = \big[ T, Y_1, Y_2, \dots, Y_{n_s} \big]$,
\item Pareto scaling, where $d_j = \sqrt{\text{std}(X_{*j})}$ with the state vector defined as $\mathbf{X} = \big[ Y_1, Y_2, \dots, Y_{n_s} \big]$.
\end{itemize}
For brevity, we refer to the dataset formulated as $\mathbf{X} = \big[ T, Y_1, Y_2, \dots, Y_{n_s} \big]$ as $\mathbf{X} = \big[ T, Y_i \big]$ and to the dataset formulated as $\mathbf{X} = \big[ Y_1, Y_2, \dots, Y_{n_s} \big]$ as $\mathbf{X} = \big[ Y_i \big]$.
The choice for the two preprocessing scenarios follows the reasoning presented in \cite{parente2013principal}, where Auto scaling was recommended for exploratory analysis and Pareto scaling was shown to align the first PC with temperature. The reason why we remove the temperature variable when using Pareto scaling in this work, is to allow for more robust linear combinations of the species mass fractions that have a better chance of serving as progress variables. We come back to this discussion in \S\ref{sec:HR}. Furthermore, when using Auto scaling, all scaled variables become dimensionless, while in Pareto scaling they do not. 

\section{Global and local PCA} \label{sec:GLPCA}

\begin{figure*}[h!]
\centering\includegraphics[width=\textwidth]{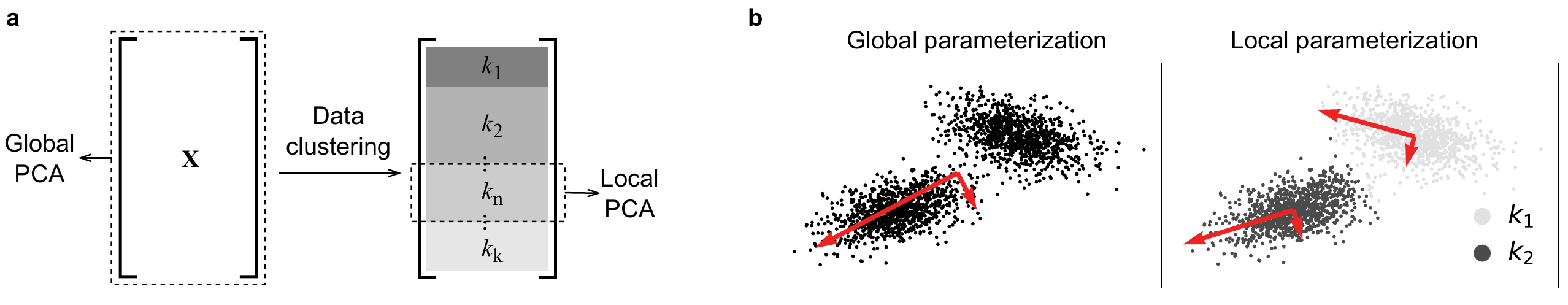}
\caption{Schematic distinction between (a) global and local PCA and (b)  global and local eigenvectors (red arrows) defining the directions of the new data parameterization. The $n^{\text{th}}$ cluster is labeled $k_n$.}
\label{fig:global-local-pca}
\end{figure*}

In this section, we briefly present the theory for global PCA, performed on the entire data matrix, $\mathbf{X}$, and for local PCA, performed in local clusters of $\mathbf{X}$, where the $n^{\text{th}}$ cluster is labeled $k_n$. The distinction between global and local PCA is schematically illustrated in Fig.~\ref{fig:global-local-pca}a.

\subsection{Global PCA} \label{sec:GPCA}

Global PCA is performed on the entire preprocessed data matrix, $\widetilde{\mathbf{X}}$. The preprocessed dataset can then be projected onto the new basis, $\mathbf{A} \in \mathbb{R}^{Q \times Q}$, such that $\mathbf{Z} = \widetilde{\mathbf{X}} \mathbf{A}$, where $\mathbf{Z} \in \mathbb{R}^{N \times Q}$ is the matrix of global PCs. The $j^{\text{th}}$ PC (the $j^{\text{th}}$ column of $\mathbf{Z}$) is thus given by
\begin{equation} \label{eq:gpca-linear-combination}
Z_{*j} = \sum_{i = 1}^Q a_{ij} X_{*i},
\end{equation}
where the coefficients of the linear combination, $a_{ij}$, are the elements of the $j^{\text{th}}$ column of $\mathbf{A}$.
PCA also provides a linear route back to the original basis,
$\mathbf{X} = \mathbf{Z} \mathbf{A}^{\top} \mathbf{D} + \mathbf{C}$
since $\mathbf{A}^{\top} = \mathbf{A}^{-1}$ for orthonormal $\mathbf{A}$.
Selecting only $q$ first PCs and eigenvectors, we can approximate the original full-rank matrix $\mathbf{X}$ by the rank-$q$ matrix $\mathbf{X_q}$, $\mathbf{X} \approx \mathbf{X_q} = \mathbf{Z_q} \mathbf{A_q}^{\top} \mathbf{D} + \mathbf{C}$, where the subscript $\mathbf{q}$ denotes truncation from $Q$ to $q$ components. The matrix of the truncated PCs is thus $\mathbf{Z_q} \in \mathbb{R}^{N \times q}$. For datasets exhibiting very strong low-rank structure, $q \ll Q$ is sufficient to accurately approximate $\mathbf{X}$ with $\mathbf{X_q}$.

\subsection{Local PCA} \label{sec:LPCA}

Once the partitioning of the dataset is obtained using any clustering algorithm of choice, PCA can be performed in local clusters of data. Let $\mathbf{X}^{(n)} \in \mathbb{R}^{N_n \times Q}$ be the observations extracted from $\mathbf{X}$ that belong to the $n^{\text{th}}$ cluster, $k_n$, with $N_n$ observations.
We also define $\widetilde{\mathbf{X}}^{(n)} \in \mathbb{R}^{N_n \times Q}$ as the analogous observations extracted from the centered and scaled global dataset, $\widetilde{\mathbf{X}}$.
Throughout this work, whenever local PCA is performed, the extracted cluster is re-centered (but not re-scaled) through $\bar{\widetilde{\mathbf{X}}}^{(n)} = \widetilde{\mathbf{X}}^{(n)}- \mathbf{C}^{(n)}$,
where $\mathbf{C}^{(n)} \in \mathbb{R}^{N_n \times Q}$ is the matrix of centers computed from $\widetilde{\mathbf{X}}^{(n)}$. The low-dimensional projection is analogous to the global PCA case, $\mathbf{Z}^{(n)} = \bar{\widetilde{\mathbf{X}}}^{(n)} \mathbf{A}^{(n)}$, where $\mathbf{Z}^{(n)} \in \mathbb{R}^{N_n \times Q}$ is the matrix of local PCs, associated with the $n^{\text{th}}$ cluster. It is worth noting, that the local eigenvectors and PCs are, in general, different from the global ones. This is illustratively demonstrated in Fig.~\ref{fig:global-local-pca}b. We refer to the $j^{\text{th}}$ local PC in the $n^{\text{th}}$ cluster as $Z_{*j}^{(n)}$ (the $j^{\text{th}}$ column of $\mathbf{Z}^{(n)}$). Finally, the local set of observations can be approximated with the rank-$q$ matrix, $\mathbf{X_q}^{(n)} $, $\mathbf{X}^{(n)} \approx \mathbf{X_q}^{(n)} = \mathbf{Z_q}^{(n)} \big(\mathbf{A_q}^{(n)} \big)^{\top} \mathbf{D} + \mathbf{C} + \mathbf{C}^{(n)}$.

\subsection{Correlation between the known and the retrieved local parameterization} \label{sec:correlation}

We utilize correlation metrics to test whether parameterization with local PCs is equivalent to the \textit{a priori} known parameterization, $\mathcal{Y}$, for datasets 1-3. Within each cluster, $k_n$, we measure correlation between the $j^{\text{th}}$ local PC, $Z_{*j}^{(n)}$, and the local parameterizing variable, denoted in general as $\mathcal{Y}^{(n)}$. We adopt two correlation metrics: Pearson correlation coefficient (PCC) and distance correlation (dCor) \citep{szekely2007measuring, edelmann2021on}.
In the $n^{\text{th}}$ cluster, the local PCC between $Z_{*j}^{(n)}$ and $\mathcal{Y}^{(n)}$ is computed as
\begin{equation}
r_n \big(Z_{*j}^{(n)}, \mathcal{Y}^{(n)} \big) = \Bigg| \frac{\sum_{i=1}^{N_n} \big(Z_{ij}^{(n)} - \overline{Z}_{*j}^{(n)} \big) \big(\mathcal{Y}_i^{(n)} - \overline{\mathcal{Y}}^{(n)} \big)}{ \Big( \sum_{i=1}^{N_n} \big(Z_{ij}^{(n)} - \overline{Z}_{*j}^{(n)} \big)^2 \sum_{i=1}^{N_n} \big(\mathcal{Y}_i^{(n)} - \overline{\mathcal{Y}}^{(n)} \big)^2 \Big)^{1/2}} \Bigg|,
\end{equation}
where $N_n$ is the total number of observations in the $n^{\text{th}}$ cluster and the overbars denote an arithmetic average over $N_n$ samples.
We further average the PCC metric over all $k$ clusters,
\begin{equation}
r(Z_{*j}, \mathcal{Y}) = \frac{1}{k} \sum_{n=1}^k r_n \big( Z_{*j}^{(n)}, \mathcal{Y}^{(n)} \big).
\end{equation}
The above equation yields a single averaged PCC value, $r$, for a given clustering solution. The $r$ values are bounded between 0 and 1. In the general case, $r=0$ is not a sufficient condition for independence between two variables.
In the $n^{\text{th}}$ cluster, the local dCor between $Z_{*j}^{(n)}$ and $\mathcal{Y}^{(n)}$ is computed as
\begin{equation} \label{eq:dCor}
\mathrm{dCor}_n \big( Z_{*j}^{(n)}, \mathcal{Y}^{(n)} \big) = \Bigg( \frac{\mathrm{dCov}_n \big( Z_{*j}^{(n)}, \mathcal{Y}^{(n)} \big)}{\mathrm{dCov}_n \big( Z_{*j}^{(n)}, Z_{*j}^{(n)} \big) \mathrm{dCov}_n \big( \mathcal{Y}^{(n)}, \mathcal{Y}^{(n)} \big)} \Bigg)^{1/2},
\end{equation}
with the local distance covariance in the $n^{\text{th}}$ cluster between two random variables, $x$ and $y$, defined as $\mathrm{dCov}_n(x,y) = \big( 1/N_n^2 \sum_{i,j=1}^{N_n} x_{ij} y_{ij} \big)^{1/2}$, where $x_{ij}$ and $y_{ij}$ are the elements of the double-centered Euclidean distances matrices for $x$ and $y$ observations respectively. The dCor metric is also averaged over $k$ clusters, 
\begin{equation}
\mathrm{dCor}(Z_{*j}, \mathcal{Y}) = \frac{1}{k} \sum_{n=1}^k \mathrm{dCor}_n \big( Z_{*j}^{(n)}, \mathcal{Y}^{(n)} \big).
\end{equation}
The $\text{dCor}$ values are bounded between 0 and 1, where $\mathrm{dCor} = 0$ iff two variables are independent.
The PCC metric can only detect linear dependence between random variables. In contrast, the dCor metric is more suitable when nonlinear association exists between variables. Thus, values of dCor larger than PCC can suggest that nonlinear dependencies remain in local portions of data. The convergence of the two metrics can indicate that the dataset has been partitioned into clusters that represent nearly linear segments of $\mathcal{Y}$.

\subsection{Clustering based on variable bins} \label{sec:clustering-mf-bins}

One way to cluster non-premixed combustion datasets is to use mixture fraction, $f$, as the conditioning variable and assign observations to bins generated in the range of $f$. This method belongs to the family of supervised clustering. The first split to two clusters can be performed at the stoichiometric value of mixture fraction, $f_{st}$, which has to be known beforehand. This first split separates observations for which $f < f_{st}$ (fuel-lean) from observations for which $f \geq f_{st}$ (fuel-rich). If the number of clusters $k > 2$, the fuel-lean and the fuel-rich branch can be further divided into segments equal in length in the $f$-space, starting from the branch which has a larger range in the values of $f$. This technique clusters the training dataset explicitly based on the chemical composition stoichiometry. For premixed systems, other physical variable, such as one of the state variables, can be selected to create the bins. In this work, whenever variable other than $f$ is used for clustering, we create $k$ equally-spaced bins between the minimum and the maximum value of that variable.

\subsection{Clustering using the VQPCA algorithm} \label{sec:clustering-vqpca}

\begin{figure}[t]
\centering\includegraphics[width=7cm]{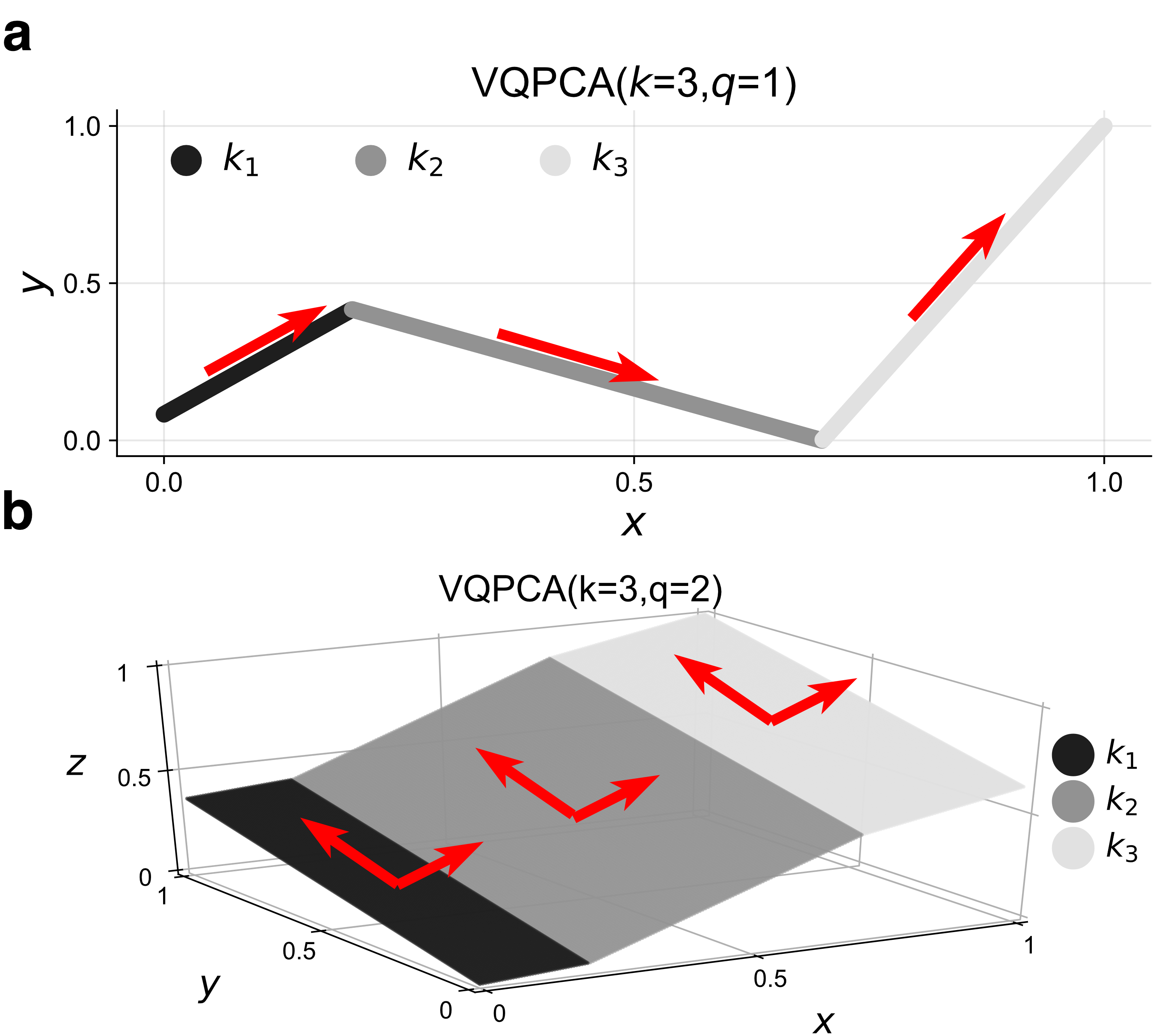}
\caption{(a) Two-dimensional and (b) three-dimensional synthetic datasets composed of hyperplanes and the result of their VQPCA clustering. Red arrows represent the expected local eigenvectors from the converged iteration, associated with each cluster. The VQPCA algorithm classifies each hyperplane as a separate cluster based on the computed eigenvectors.}
\label{fig:vqpca-synthetic-data}
\end{figure}

VQPCA is an iterative clustering algorithm that partitions the data based on the minimum reconstruction error from local PCA \citep{kambhatla1997dimension, parente2009identification, zdybal2022advancing}. Performing clustering through dimensionality reduction is a unique feature of VQPCA. Compared to other unsupervised techniques such as K-Means clustering, VQPCA has a potential advantage of being less sensitive to noise or outliers in the data. The convergence of the algorithm is assessed with: (1) the convergence of cluster centroids (the centroid of each cluster should not move more than a threshold between two consecutive iterations) and (2) the convergence of the PCA reconstruction error (the error should be smaller than an agreed threshold).
The algorithm takes the preprocessed dataset, $\widetilde{\mathbf{X}}$, and the following hyper-parameters as inputs:
\begin{itemize}
\item the number of clusters, $k$, to create in the dataset,
\item the number of local eigenvectors, $q$, with which each observation is approximated at each iteration, and
\item the initialization of cluster centroids, $v_0 \in \mathbb{Z}_+^{N\times1}$, defining the initial cluster classifications.
\end{itemize}
For simplicity, we refer to the VQPCA solution with $k$ clusters and $q$ local eigenvectors as VQPCA$(k,q)$. The VQPCA algorithm is bootstrapped with the initial cluster centroids, $v_0$, which can be obtained using any data partitioning technique of choice. The most straightforward way is to randomly assign original observations into $k$ clusters. Another option is to uniformly stratify the data matrix into initial clusters. In this work, we use bins of mixture fraction as the cluster initializations on datasets BS, EQ, DNS and SLF and equally-spaced bins of \ch{H2O} mass fractions for the HR dataset (as described in \S\ref{sec:clustering-mf-bins}). 

At each iteration of the VQPCA algorithm, the dataset is approximated using a separate set of local eigenvectors computed from each of the $k$ clusters. A matrix of reconstruction errors, $\pmb{\varepsilon}\in \mathbb{R}^{N \times k}$, is populated, such that each column is linked to one of the $k$ clusters. Each $(i,j)$-th element, $\varepsilon_{ij}$, of that matrix represents the reconstruction error of the $i^{\text{th}}$ observation if it was assigned to the $j^{\text{th}}$ cluster.
The error is computed based on the global scaled reconstruction error metric as per \cite{parente2009identification},
\begin{equation} \label{eq:vqpca-rec-error}
\varepsilon_{ij} = || \widetilde{X}_{i*} - \widetilde{X}_{q, i*}^{(j)}||,
\end{equation}
where $\widetilde{X}_{i*}$ is the preprocessed $i^{\text{th}}$ observation of all $Q$ variables in the dataset and $\widetilde{X}_{q, i*}^{(j)}$ is the rank-$q$ approximation of that observation using local eigenvectors from the $j^{\text{th}}$ cluster. The minimum reconstruction error is then searched across every row of $\pmb{\varepsilon}$. An individual observation is eventually assigned to the cluster in which that error was the smallest. This allows to form a new vector of cluster classifications and re-compute the local eigenvectors to be used at the next iteration. If the number of observations assigned to a particular cluster is less than the number of variables in a dataset, the cluster is removed. The observations from that removed cluster are redistributed at the next iteration. This prevents forming spurious clusters. Once the VQPCA algorithm converges, we obtain the final vector defining cluster classifications, $v \in \mathbb{Z}_+^{N\times1}$.

\begin{figure}[t]
\centering\includegraphics[width=7cm]{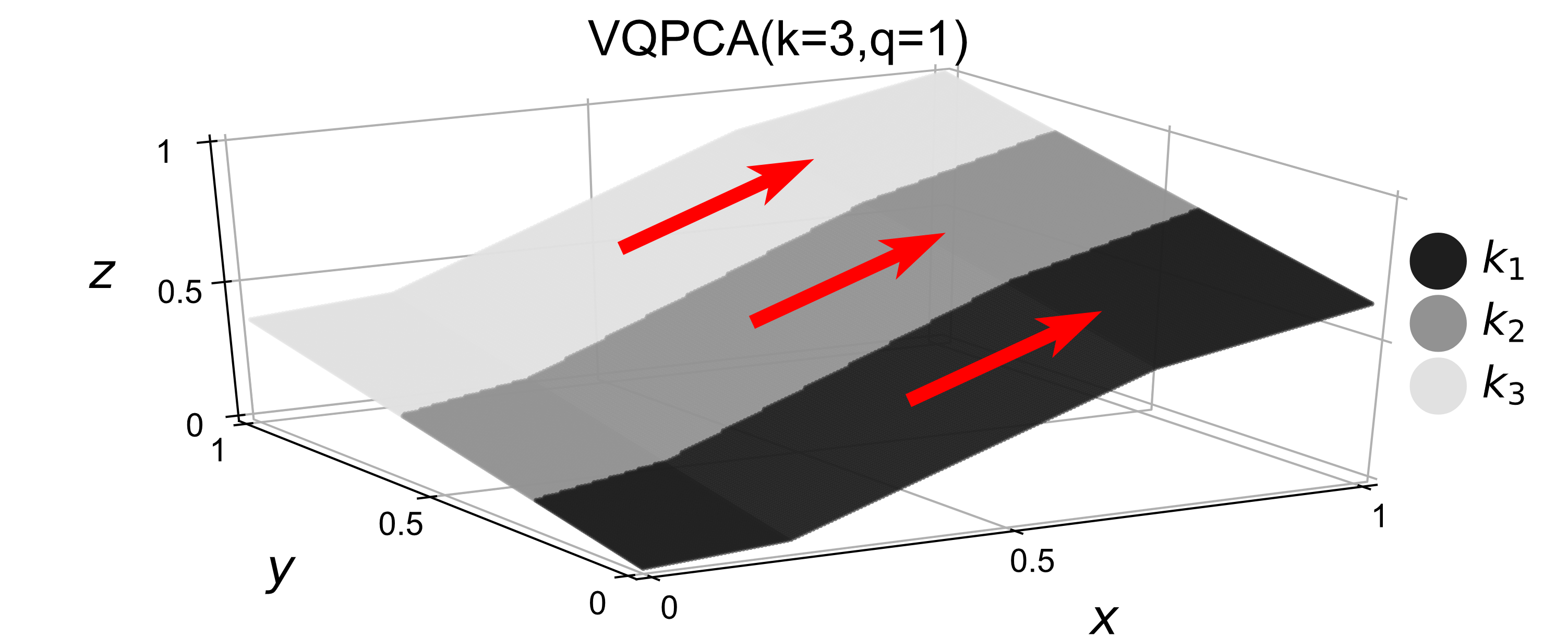}
\caption{The result of clustering the synthetic dataset from Fig.~\ref{fig:vqpca-synthetic-data}b using VQPCA($k$=3,$q$=1). Red arrows represent the possible local eigenvectors from the converged iteration, associated with each cluster.}
\label{fig:vqpca-synthetic-data-failed-q-guess}
\end{figure}

In essence, each cluster obtained with the VQPCA algorithm is associated with some $q$-dimensional subspace of the original $Q$-dimensional space. We define linearity in $q$ dimensions as observations that lie within the span of a $q$-dimensional orthogonal basis. In a special case, when the dataset is composed of $k$ $q$-dimensional linear manifolds, we can expect that VQPCA$(k,q)$ will classify each manifold as a separate cluster. In a more general case, for datasets that are not composed of $q$-dimensional manifolds or, when we fail to guess how many underlying manifolds there are, each obtained cluster lies close to a $q$-dimensional manifold.  The ``closeness'' is measured by the reconstruction errors as per Eq.~(\ref{eq:vqpca-rec-error}). Fig.~\ref{fig:vqpca-synthetic-data} visualizes this concept on two synthetic datasets composed of hyperplanes. The first one, in Fig.~\ref{fig:vqpca-synthetic-data}a, is composed of three line segments embedded in a two-dimensional space. When VQPCA($k$=3,$q$=1) is applied to this data, a single eigenvector is allowed to define each of the three clusters. Since the span of a single vector can define a line, the expectation is that VQPCA will partition this dataset such that each line segment becomes one cluster. The second dataset, in Fig.~\ref{fig:vqpca-synthetic-data}b, is composed of three planar segments embedded in a three-dimensional space. This time, when VQPCA($k$=3,$q$=2) is applied, we allow two orthogonal eigenvectors to be associated with each cluster. This enables detecting planar structures and, in an analogous way to Fig.~\ref{fig:vqpca-synthetic-data}a, we can expect VQPCA to assign each of the three planes to a separate cluster. We observe that the algorithm performed as expected; each hyperplane is classified as a separate cluster based on the computed eigenvectors. The expected eigenvectors from the converged iteration are schematically illustrated with red arrows.

A potential shortcoming of the VQPCA algorithm emerges following our discussion. VQPCA cannot guarantee a ``reasonable'' clustering solution if one fails to guess the dimensionality, $q$, of the underlying manifolds. This is demonstrated in Fig.~\ref{fig:vqpca-synthetic-data-failed-q-guess}, where we apply VQPCA($k$=3,$q$=1) on the synthetic dataset from Fig.~\ref{fig:vqpca-synthetic-data}b. Constraining VQPCA to fit one-dimensional manifolds ($q$=1) results in a different clustering solution from the one presented in Fig.~\ref{fig:vqpca-synthetic-data}b. Unsurprisingly, the existence of the three planar segments is now ignored. The clustering result still satisfies optimality from the perspective of the reconstruction error metric (as per Eq.~(\ref{eq:vqpca-rec-error})). Hence, in Fig.~\ref{fig:vqpca-synthetic-data-failed-q-guess}, we see three elongated clusters, each being a result of assigning observations to the span of a single local eigenvector (schematically visualized with the red arrows). Comparing clustering results in Figs.~\ref{fig:vqpca-synthetic-data}b and \ref{fig:vqpca-synthetic-data-failed-q-guess}, it becomes evident that changing the hyper-parameter $q$ can result in very different clusters. 

\begin{figure}[b]
\centering\includegraphics[width=7cm]{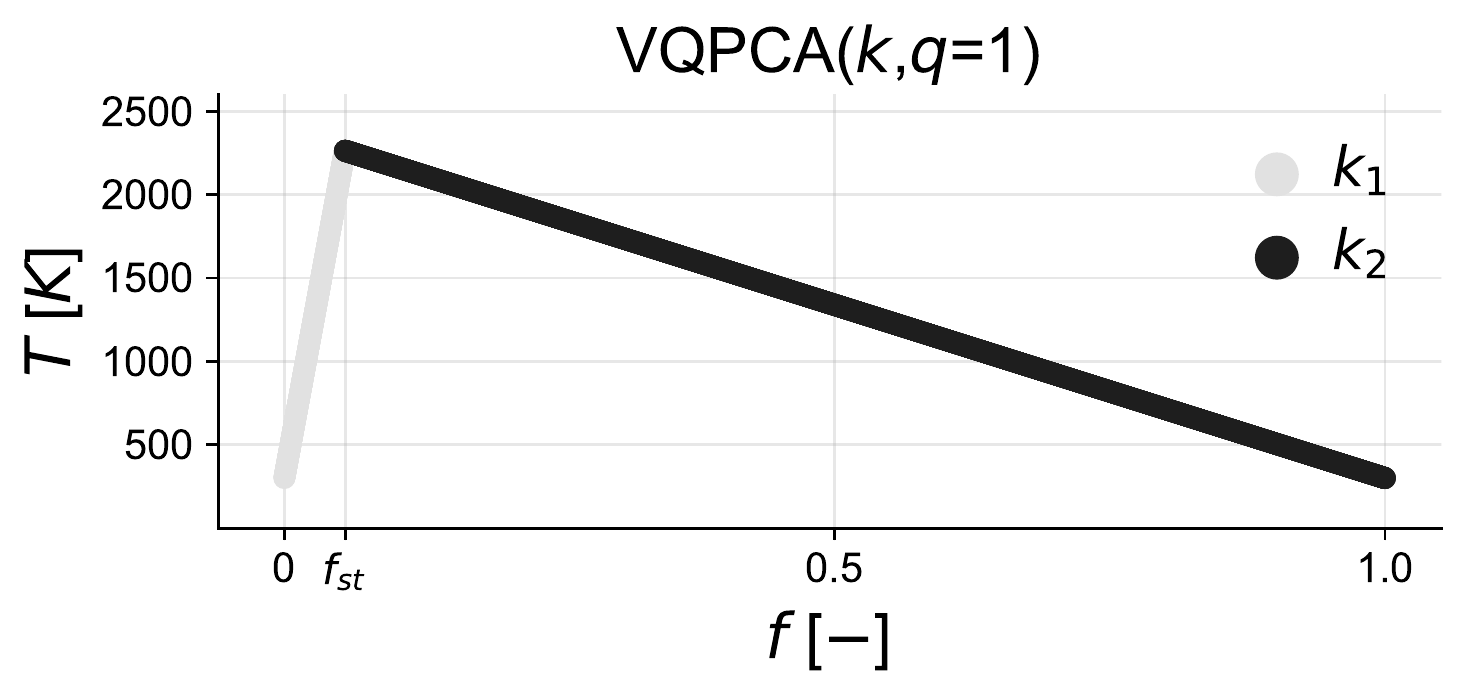}
\caption{The result of VQPCA clustering of the BS dataset using $q$=1 (irrespective of the number of clusters, $k$, originally requested). $f_{st}$ denotes the stoichiometric value of $f$. Local PCA was performed using Auto scaling on a dataset $\mathbf{X} = \big[T, Y_i \big]$. An identical result is obtained when local PCA is performed using Pareto scaling on a dataset $\mathbf{X} = \big[Y_i \big]$. Clusters were initialized by binning $f$.}
\label{fig:BS-clustering}
\end{figure}

This shortcoming calls for measures that allow for automatic detection of  the intrinsic dimensionality of the local manifolds. While several metrics exist for approximating the effective or intrinsic dimensionality \cite{del2021effective}, this task remains somewhat challenging. One way in which we can guide the choice for $q$ in an \textit{a posteriori} analysis of clustering solutions is by measuring the average reconstruction accuracy of each cluster using the local eigenvectors associated with that cluster. For example, the coefficient of determination, $R^2$, is equal to unity when the three clusters from the VQPCA($k$=3,$q$=2) solution in Fig.~\ref{fig:vqpca-synthetic-data}b are reconstructed. For the VQPCA($k$=3,$q$=1) solution, the average $R^2$ for the reconstruction of the three clusters seen in Fig.~\ref{fig:vqpca-synthetic-data-failed-q-guess} is 0.69. The supplementary material presents extended results for this synthetic dataset (see Fig.~S1) for other combinations of $k$ and $q$. A non-unity value of $R^2$ conveys the fact that observations in a cluster have some level of scatter around the span of their defining eigenvector(s). Hence, observations cannot be reconstructed perfectly and, at best, lie close to that span. For datasets such as the DNS data, this is generally the case due to large level of physical scatter in the data. Tied to this discussion, although perhaps less cumbersome, is the need to find the optimal number of clusters, $k$, that sufficiently represent distinct physical phenomena. This point can be addressed using clustering quality metrics that measure intra-cluster homogeneity and inter-cluster heterogeneity. We come back to this discussion in \S\ref{sec:DNS} and present an approach for a more informed selection of the pair $(k,q)$ tied to a given dataset, when no \textit{a priori} physical knowledge is available to guide this choice.

\section{Results and discussion}

\subsection{The Burke-Schumann model} \label{sec:BS}

The BS dataset carries similarities with the synthetic dataset presented in Fig.~\ref{fig:vqpca-synthetic-data}a; by construction, this dataset is parameterized by $\mathcal{Y} = f$ only. Since the lean and rich branch are linear functions of $f$, clustering with VQPCA($k$=2,$q$=1) should separate the lean and rich branch into two clusters. The result of this clustering is presented in Fig.~\ref{fig:BS-clustering}. VQPCA splits the dataset at the stoichiometric mixture fraction value, $f_{st}$, and the two clusters correspond to the fuel-lean and the fuel-rich side. Such partition leads to 100\% correlation in terms of $r$ and dCor between the first local PC and the local $f$ in both clusters. Further, we have found that no matter how many clusters, $k$, are requested for partitioning this dataset, VQPCA ultimately finds only two, removing spurious clusters that have too few observations. This behavior is understandable, since the BS dataset is composed of only two locally one-dimensional manifolds. It is important to note here, that this finding does not prove that our local PCA approach ``sees'' the underlying physical meaning of the two clusters \textit{a priori}. Rather, retrieving the two clusters is an aftermath of the mechanics of the VQPCA algorithm (see \S\ref{sec:clustering-vqpca}). Since the BS model is constructed using two one-dimensional linear manifolds, the one-dimensional manifolds found by VQPCA concur entirely with our physical interpretation of these manifolds.

\subsection{The chemical equilibrium model} \label{sec:EQ} 

\begin{figure*}[h!]
\centering\includegraphics[width=\textwidth]{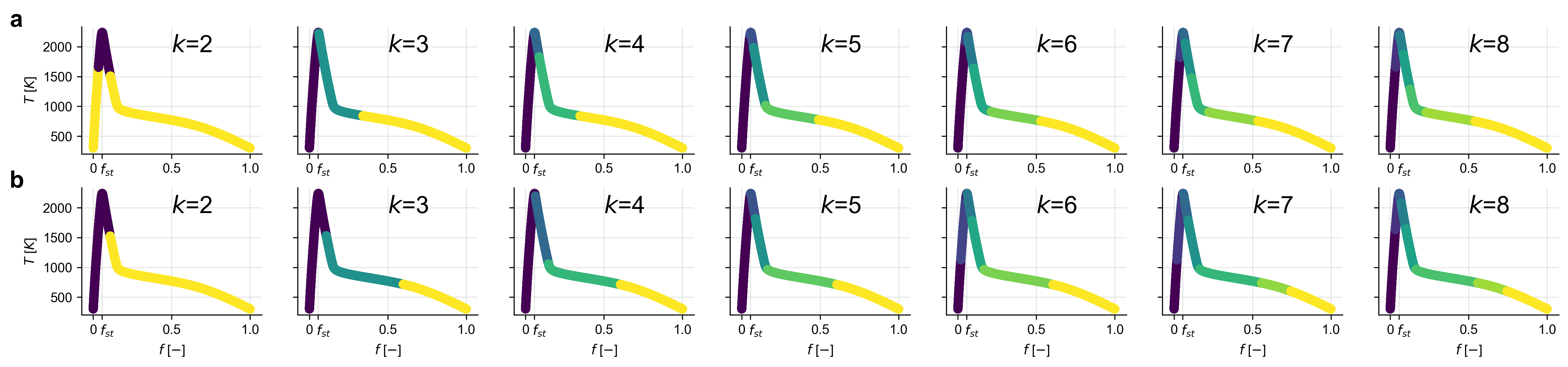}
\caption{The result of VQPCA clustering of the EQ dataset to $k$ clusters using $q$=1. $f_{st}$ denotes the stoichiometric value of $f$.
(a) Local PCA was performed using Auto scaling on a dataset $\mathbf{X} = \big[T, Y_i \big]$.
(b) Local PCA was performed using Pareto scaling on a dataset $\mathbf{X} = \big[Y_i \big]$. 
In (a) and (b), clusters were initialized by binning $f$.}
\label{fig:EQ-clustering}
\end{figure*}

\begin{figure}[t]
\centering\includegraphics[width=7cm]{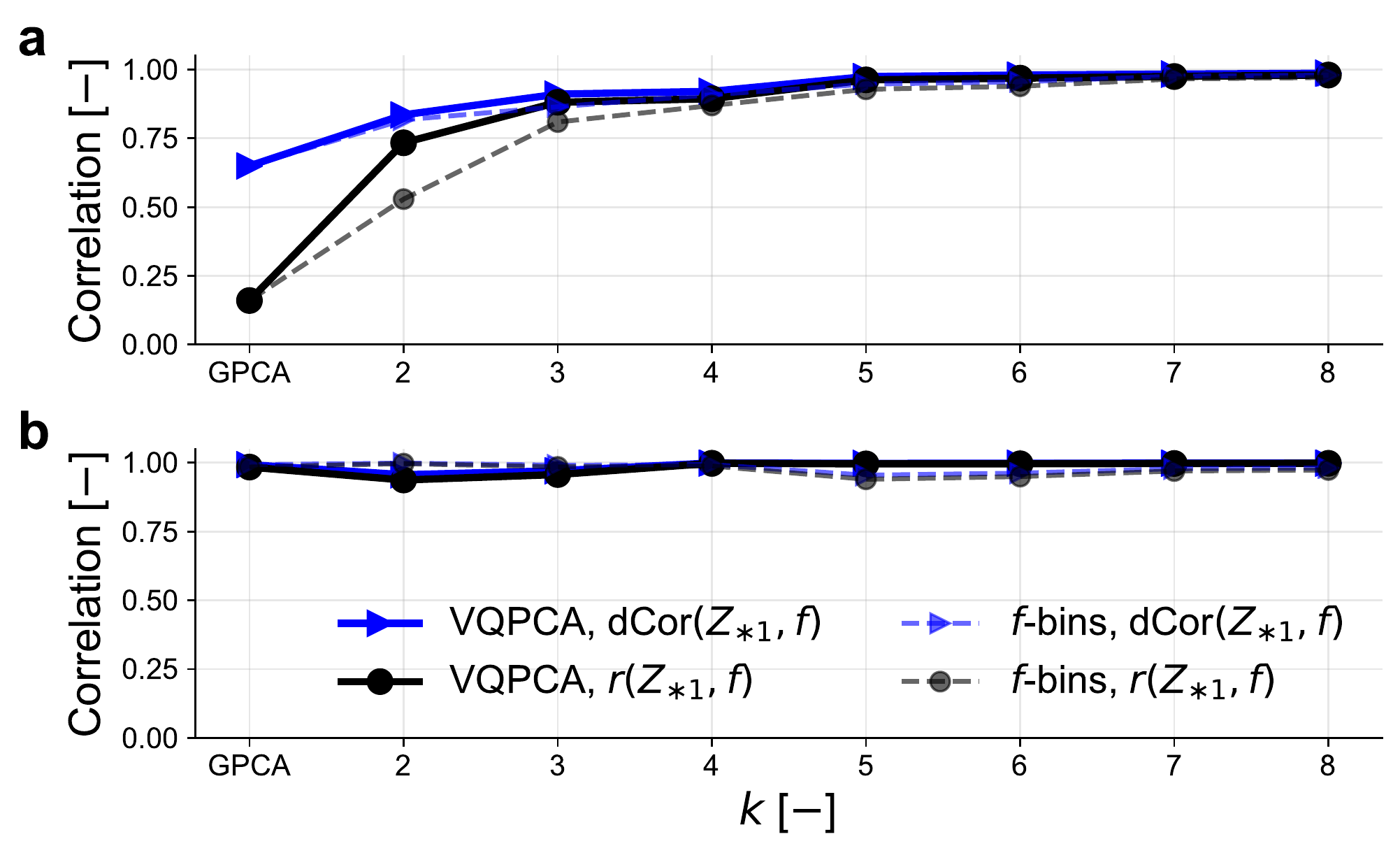}
\caption{The $r$ and dCor values between the first local PC and the local $\mathcal{Y} = f$ when clustering the EQ dataset with VQPCA with an increasing number of clusters, $k$, and using $q$=1. For comparison, dashed lines show analogous correlations but for clustering solutions obtained directly from binning the $f$ vector. We also show correlations coming from global PCA (marked with GPCA).
(a) Global and local PCA were performed using Auto scaling on a dataset $\mathbf{X} = \big[T, Y_i \big]$.
(b) Global and local PCA were performed using Pareto scaling on a dataset $\mathbf{X} = \big[Y_i \big]$. 
In (a) and (b), clusters were initialized by binning $f$. Legend applies to (a) and (b).}
\label{fig:EQ-clustering-correlations}
\end{figure}

\begin{figure*}[t]
\centering\includegraphics[width=\textwidth]{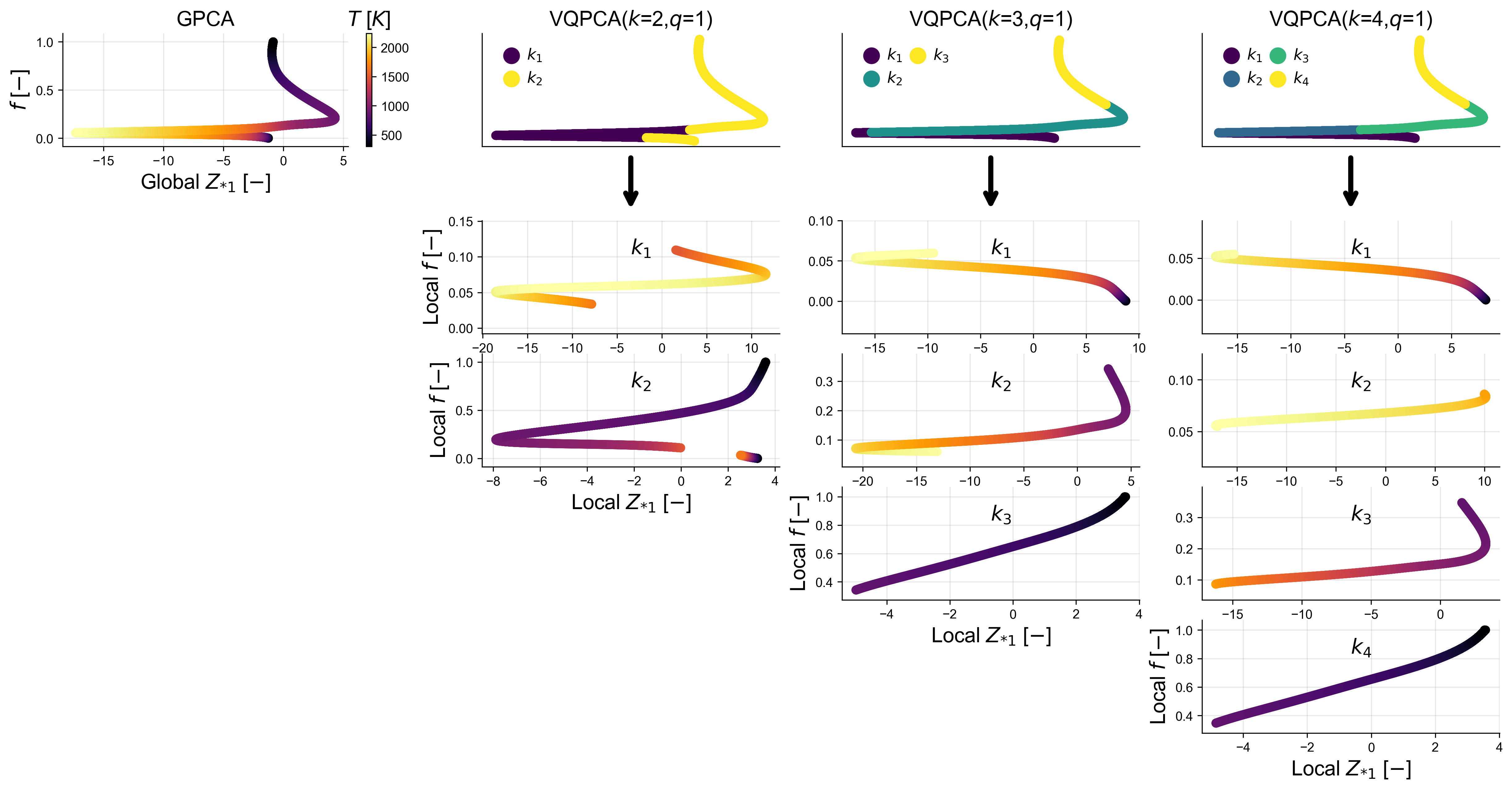}
\caption{True shapes of the global and local one-dimensional manifolds in the EQ dataset as functions of $f$. We report the one-dimensional manifolds found by global PCA (GPCA) and local PCA when $k$ was varied from 2 to 4 in VQPCA($k$,$q$=1). Local PCA was performed using Auto scaling on a dataset $\mathbf{X} = \big[T, Y_i \big]$. Clusters were initialized by binning $f$. The same temperature colorbar applies to global and local manifolds.}
\label{fig:EQ-true-local-manifolds-auto}
\end{figure*}

\begin{figure*}[t]
\centering\includegraphics[width=\textwidth]{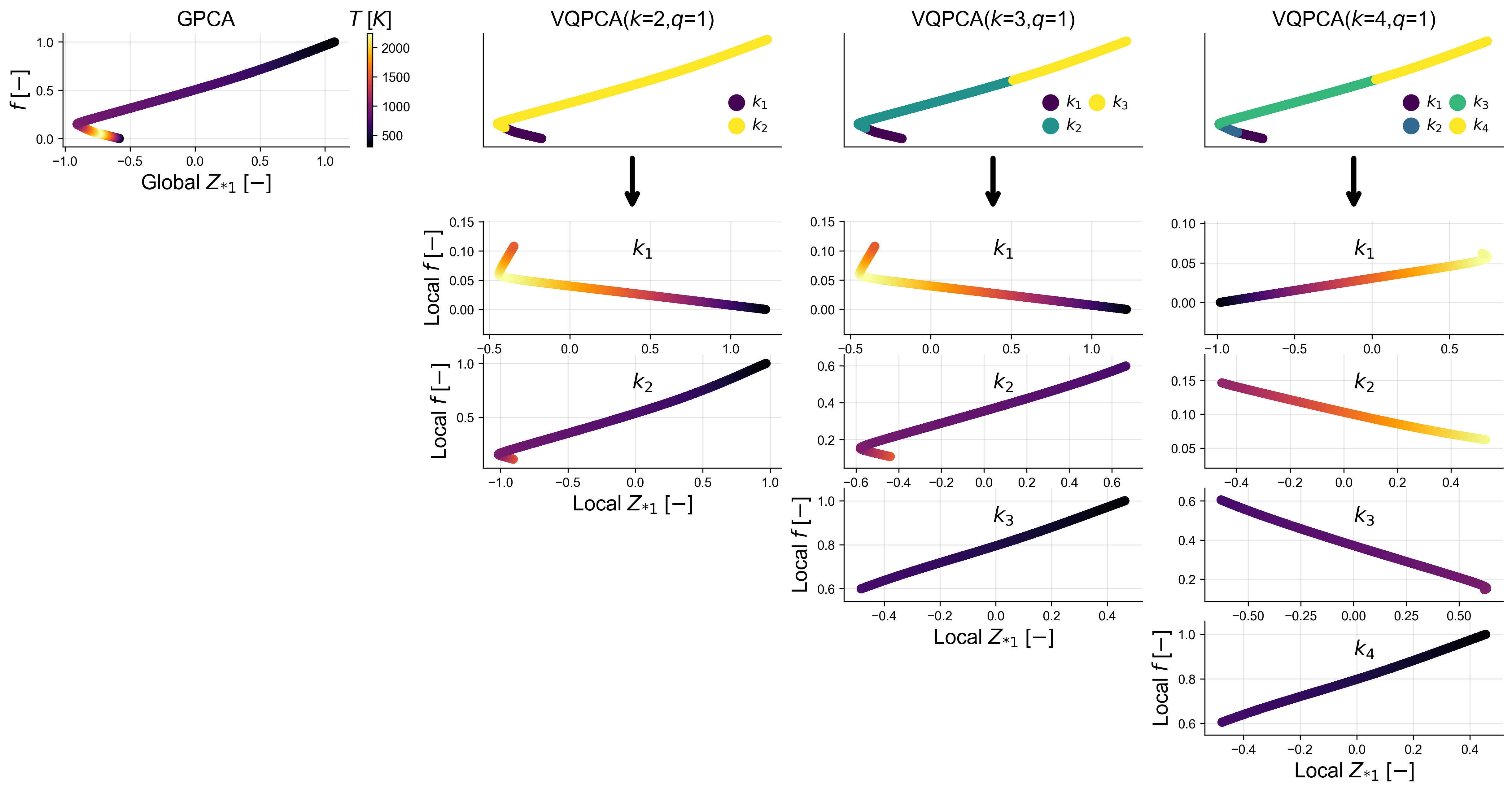}
\caption{True shapes of the global and local one-dimensional manifolds in the EQ dataset as functions of $f$. We report the one-dimensional manifolds found by global PCA (GPCA) and local PCA when $k$ was varied from 2 to 4 in VQPCA($k$,$q$=1). Local PCA was performed using Pareto scaling on a dataset $\mathbf{X} = \big[Y_i \big]$. Clusters were initialized by binning $f$. The same temperature colorbar applies to global and local manifolds.}
\label{fig:EQ-true-local-manifolds-pareto}
\end{figure*}

Similarly to the BS dataset, the EQ dataset is inherently parameterized by $\mathcal{Y} = f$ but the local manifolds are not linear. Fig.~\ref{fig:EQ-clustering} shows the result of VQPCA clustering of the EQ dataset to $k$ clusters using $q$=1, where $k$ varies from 2 to 8. We compare two data preprocessing scenarios. In Fig.~\ref{fig:EQ-clustering}a, local PCA was performed using Auto scaling on a dataset $\mathbf{X} = \big[T, Y_i \big]$. In Fig.~\ref{fig:EQ-clustering}b, local PCA was performed using Pareto scaling on a dataset $\mathbf{X} = \big[Y_i \big]$. In both cases, clusters were initialized by binning the $f$ vector. In Fig.~\ref{fig:EQ-clustering}a, when creating two clusters, the first cluster captures all observations from the pure streams. In the space of $f$-$T$, this first cluster is discontinuous. It is worth noting, however, that $f$ does not constitute the dimensions of the ambient space. The fact that VQPCA classified the observations in the oxidizer and fuel stream into one cluster means that those observations lie sufficiently close to the same local manifold, embedded in the ambient space. Starting from $k$=3 clusters, VQPCA uses the additional third eigenvector to construct separate manifolds for the oxidizer and fuel stream. It is visible in Fig.~\ref{fig:EQ-clustering}a that also starting from $k$=3, VQPCA splits the dataset close to $f_{st}$. In the space of $f$-$T$, this results in splitting the dataset at the location of the largest visible nonlinearity. For the data preprocessing case from Fig.~\ref{fig:EQ-clustering}b, we do not observe discontinuous clusters. However, the split close to $f_{st}$ happens only starting from $k$=4.

In order to interpret the one-dimensional local manifolds that VQPCA finds for this dataset, we compute the $r$ and dCor values between the first local PC and the local $\mathcal{Y} = f$, for each clustering solution from Fig.~\ref{fig:EQ-clustering}. These values, for increasing $k$, are reported in Fig.~\ref{fig:EQ-clustering-correlations} with solid lines. We obtain a converging behavior in the correlation metrics as the number of clusters increases. For Fig.~\ref{fig:EQ-clustering-correlations}a, which corresponds to the preprocessing case with Auto scaling on a dataset $\mathbf{X} = \big[T, Y_i \big]$, the minimum correlation in terms of $r$ is 73\% when $k$=2. For Fig.~\ref{fig:EQ-clustering-correlations}b, which corresponds to the preprocessing case with Pareto scaling on a dataset $\mathbf{X} = \big[Y_i \big]$, the minimum correlation in terms of $r$ is 94\% when $k$=2. Thus, there is a good agreement between the parameterization obtained with local PCA and the known intrinsic parameterization of this EQ dataset ($\mathcal{Y} = f$).

We note, that even though clusters were initialized from binning the $f$ vector, VQPCA re-distributes observations during the iterative process with respect to the initialized solution. The re-distribution is such that correlations are generally higher for the final VQPCA clustering solution as compared to the clustering solution coming from binning $f$ alone (cf. solid and dashed lines in Fig.~\ref{fig:EQ-clustering-correlations}). While we have found binning the $f$ vector to be a good choice for initializing clusters in the VQPCA algorithm, the final VQPCA clustering solution is improved with respect to simply binning the $f$ vector in terms of retrieving the locally one-dimensional manifolds. 

\begin{figure*}[t]
\centering\includegraphics[width=\textwidth]{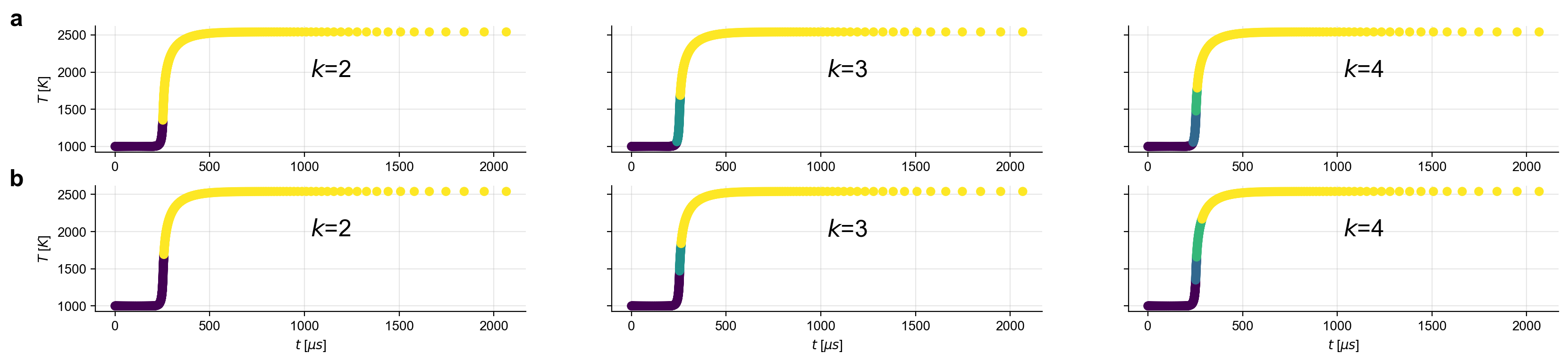}
\caption{The result of VQPCA clustering of the HR dataset to $k$ clusters using $q=1$.
(a) Local PCA was performed using Auto scaling on a dataset $\mathbf{X} = \big[T, Y_i \big]$.
(b) Local PCA was performed using Pareto scaling on a dataset $\mathbf{X} = \big[Y_i \big]$. 
In (a) and (b), clusters were initialized by binning $Y_{H_2O}$.}
\label{fig:HR-clustering}
\end{figure*}

There is a visible larger discrepancy between $r$ and dCor values when $k<5$, especially for the case of Auto scaling on a dataset $\mathbf{X} = \big[T, Y_i \big]$. This suggests that for small $k$, local manifolds still exhibit some level of nonlinearity. It is thus instructive to inspect the true shapes of the local one-dimensional manifolds in functions of $\mathcal{Y} = f$ for the EQ dataset. Fig.~\ref{fig:EQ-true-local-manifolds-auto} shows these true shapes for global PCA and for local PCA performed in clusters found by VQPCA where $k$ varies from 2 to 4. We note that the global manifold (denoted GPCA) and the local manifolds associated with VQPCA($k$=2,$q$=1), exhibit significant nonlinearity. For the VQPCA($k$=3,$q$=1) solution, we obtain at least one local manifold (in cluster $k_3$) which is nearly a linear function of $f$. For the VQPCA($k$=4,$q$=1) solution, the linearity of local manifolds is more pronounced, with manifolds in at least three clusters ($k_1$, $k_2$ and $k_4$) being nearly piecewise linear one-dimensional manifolds.
Fig.~\ref{fig:EQ-true-local-manifolds-pareto} shows analogous true shapes of the local one-dimensional manifolds in the EQ dataset, but for the preprocessing case using Pareto scaling on a dataset $\mathbf{X} = \big[Y_i \big]$. We observe that the one-dimensional manifold from global PCA is already close to being a linear in function of $f$, as opposed to the global PCA manifold seen in Fig.~\ref{fig:EQ-true-local-manifolds-auto}. With local PCA performed in VQPCA clusters, we obtain nearly linear one-dimensional manifolds, especially when creating four clusters (the VQPCA($k$=4,$q$=1) solution). This qualitative comparison of local manifolds in Figs.~\ref{fig:EQ-true-local-manifolds-auto}-\ref{fig:EQ-true-local-manifolds-pareto} between the two preprocessing scenarios can help explain the generally much higher (nearing 100\%) correlations seen in Fig.~\ref{fig:EQ-clustering-correlations}b (for Pareto scaling with $\mathbf{X} = \big[Y_i \big]$), as compared to Fig.~\ref{fig:EQ-clustering-correlations}a (for Auto scaling with $\mathbf{X} = \big[T, Y_i \big]$). This also demonstrates that even for qualitatively similar clustering solutions (cf. Figs.~\ref{fig:EQ-clustering}a-b), data preprocessing can play a significant role in local manifold parameterization and interpretation. Some means of data preprocessing (such as Pareto scaling applied on $\mathbf{X} = \big[Y_i \big]$) can help align manifold parameterizations with the physically expected parameterization of the system.
The supplementary material presents analogous results for hydrogen/air and syngas/air combustion (see Figs.~S2-S9).

\subsection{The homogeneous reactor model} \label{sec:HR}

\begin{figure}[t]
\centering\includegraphics[width=7cm]{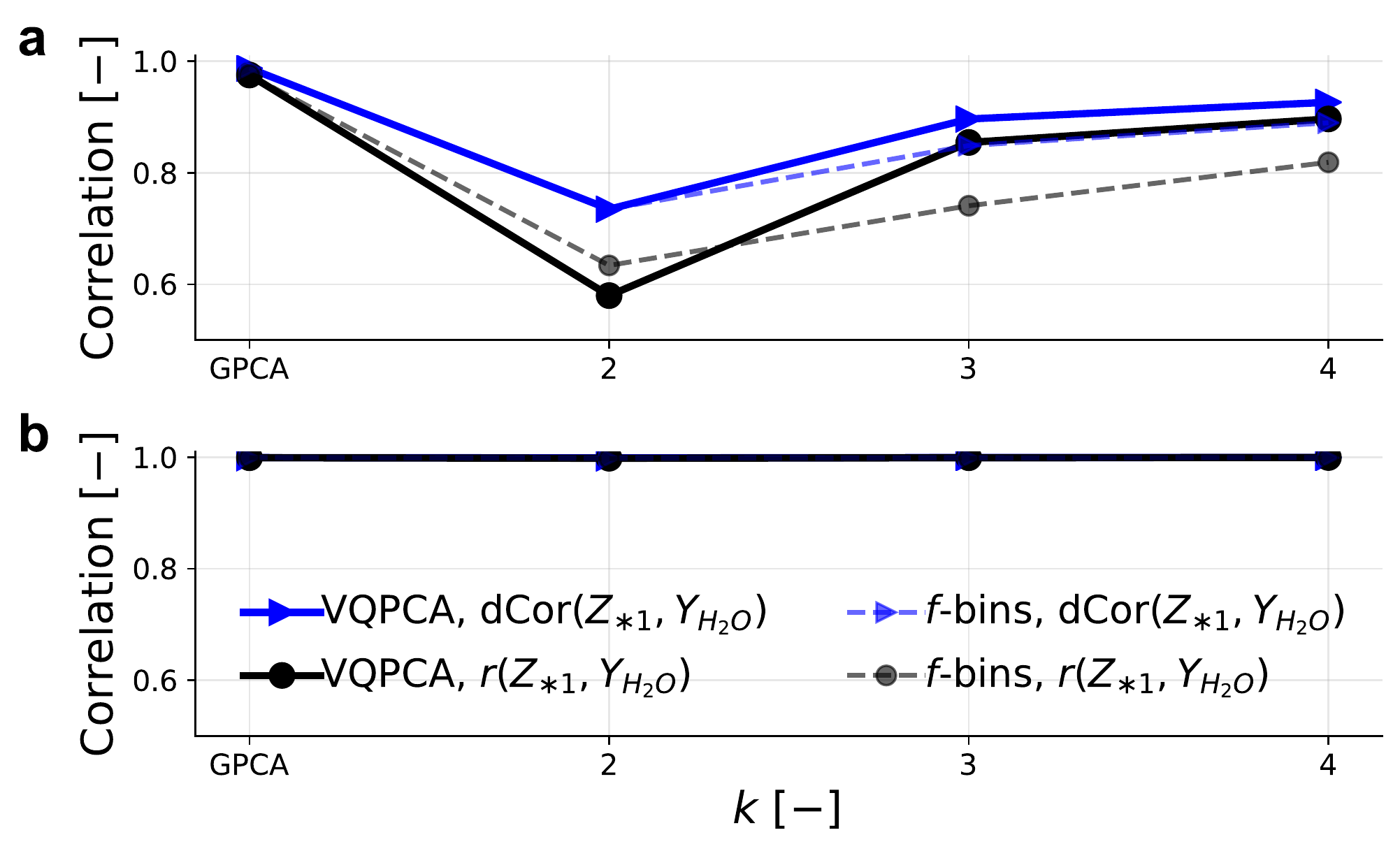}
\caption{The $r$ and dCor values between the first local PC and the local $\mathcal{Y} = Y_{H_2O}$ when clustering the HR dataset with VQPCA with an increasing number of clusters, $k$, and using $q$=1. For comparison, dashed lines show analogous correlations but for clustering solutions obtained directly from binning the $Y_{H_2O}$ vector. We also show correlations coming from global PCA (marked with GPCA).
(a) Local PCA was performed using Auto scaling on a dataset $\mathbf{X} = \big[T, Y_i \big]$.
(b) Local PCA was performed using Pareto scaling on a dataset $\mathbf{X} = \big[Y_i \big]$.
In (a) and (b), clusters were initialized by binning $Y_{H_2O}$. Legend applies to (a) and (b).}
\label{fig:HR-clustering-correlations}
\end{figure}

We apply the local PCA approach to a single time trajectory of the HR dataset. The HR data results from solving a system of ODEs with physical time, $t$, being the independent variable. For this dataset, time is the variable describing the reaction progress, from initial condition in the reactor, through mixture ignition to chemical equilibrium. Fig.~\ref{fig:HR-clustering} shows the result of VQPCA clustering of the HR dataset to $k$ clusters, in the space of time, $t$, (the independent variable) and temperature, $T$. Here, $k$ varies only from 2 to 4 due to small number of observations in the HR dataset. For $k>4$, some clusters become too small to be considered meaningful for the correlation analysis. Since a single time trajectory is used, we select $q$=1 in the VQPCA algorithm. Since the HR dataset describes a premixed system, VQPCA is now initialized with equally-spaced bins of the \ch{H2O} mass fraction, $Y_{H_2O}$. We observe that clusters are continuous along the time dimension. This suggests that the VQPCA algorithm clusters the HR dataset based on a variable describing progress of chemical reactions.

Mass fractions of combustion products, (in the case of hydrogen fuel, \ch{H2O}) can be used to define progress variables \citep{chen2004experimental, fiolitakis2010novel, ladeinde2018differential, vasavan2020novel}. In general, a progress variable, $c$, is defined in the literature as $c = \sum_{i=1}^{n_s} w_i Y_i$, where the weighting factors, $w_i$, are typically non-zero only for stable species and such that their linear combination changes monotonically across the flame. There is a similarity of this definition with Eq.~(\ref{eq:gpca-linear-combination}), which suggests that under certain settings, a PC can satisfy the role of a progress variable. The dataset for PCA transformation needs to be composed of species mass fractions only (i.e. the temperature variable is not included in the training data). In Fig.~\ref{fig:HR-clustering-correlations}, with solid lines we show correlations between the local PC in each cluster and the \ch{H2O} mass fraction ($\mathcal{Y} = Y_{H_2O}$). For Fig.~\ref{fig:HR-clustering-correlations}a, which corresponds to the preprocessing case with Auto scaling on a dataset $\mathbf{X} = \big[T, Y_i \big]$, the minimum correlation in terms of $r$ is 58\% when $k$=2. For Fig.~\ref{fig:HR-clustering-correlations}b, which corresponds to the preprocessing case with Pareto scaling on a dataset $\mathbf{X} = \big[Y_i \big]$, the correlation in terms of $r$ is 100\% for all $k$ considered. We thus see consistency of the local PCA results with the clusters visualization presented in Fig.~\ref{fig:HR-clustering}. The VQPCA algorithm partitioned the dataset based on parameters highly correlated with a progress variable. This is especially the case when Pareto scaling was used on a dataset with the temperature variable removed, following our earlier expectation about how $c$ is formed. In the supplement, we show analogous results for syngas/air and methane/air combustion (see Figs.~S10-S13). We also show correlations between the first local PC and $\mathcal{Y} = Y_{CO_2}$, which are of relevance for hydrocarbon fuels (see Figs.~S14-S15).

\begin{figure}[t]
\centering\includegraphics[width=8cm]{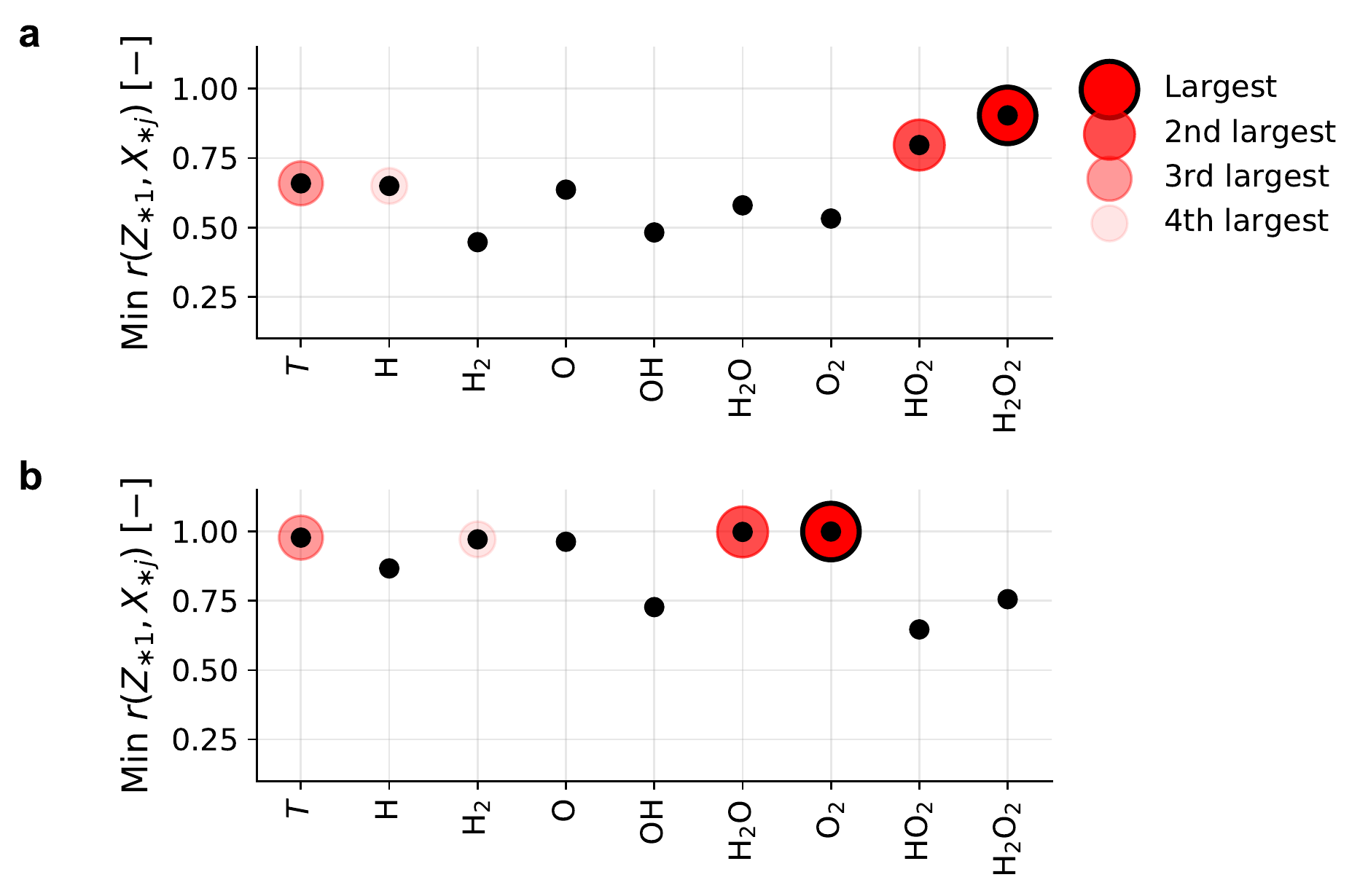}
\caption{The minimum $r$ values between the first local PC and each local state variable, $X_{*j}$, when clustering the HR dataset for \textbf{hydrogen/air} combustion with VQPCA with an increasing number of clusters, $k$, and using $q$=1.
(a) Local PCA was performed using Auto scaling on a dataset $\mathbf{X} = \big[T, Y_i \big]$.
(b) Local PCA was performed using Pareto scaling on a dataset $\mathbf{X} = \big[Y_i \big]$. 
In (a) and (b), clusters were initialized by binning $Y_{H_2O}$. Legend applies to (a) and (b).}
\label{fig:HR-clustering-minimum-correlations-H2}
\end{figure}

\begin{figure}[t]
\centering\includegraphics[width=8cm]{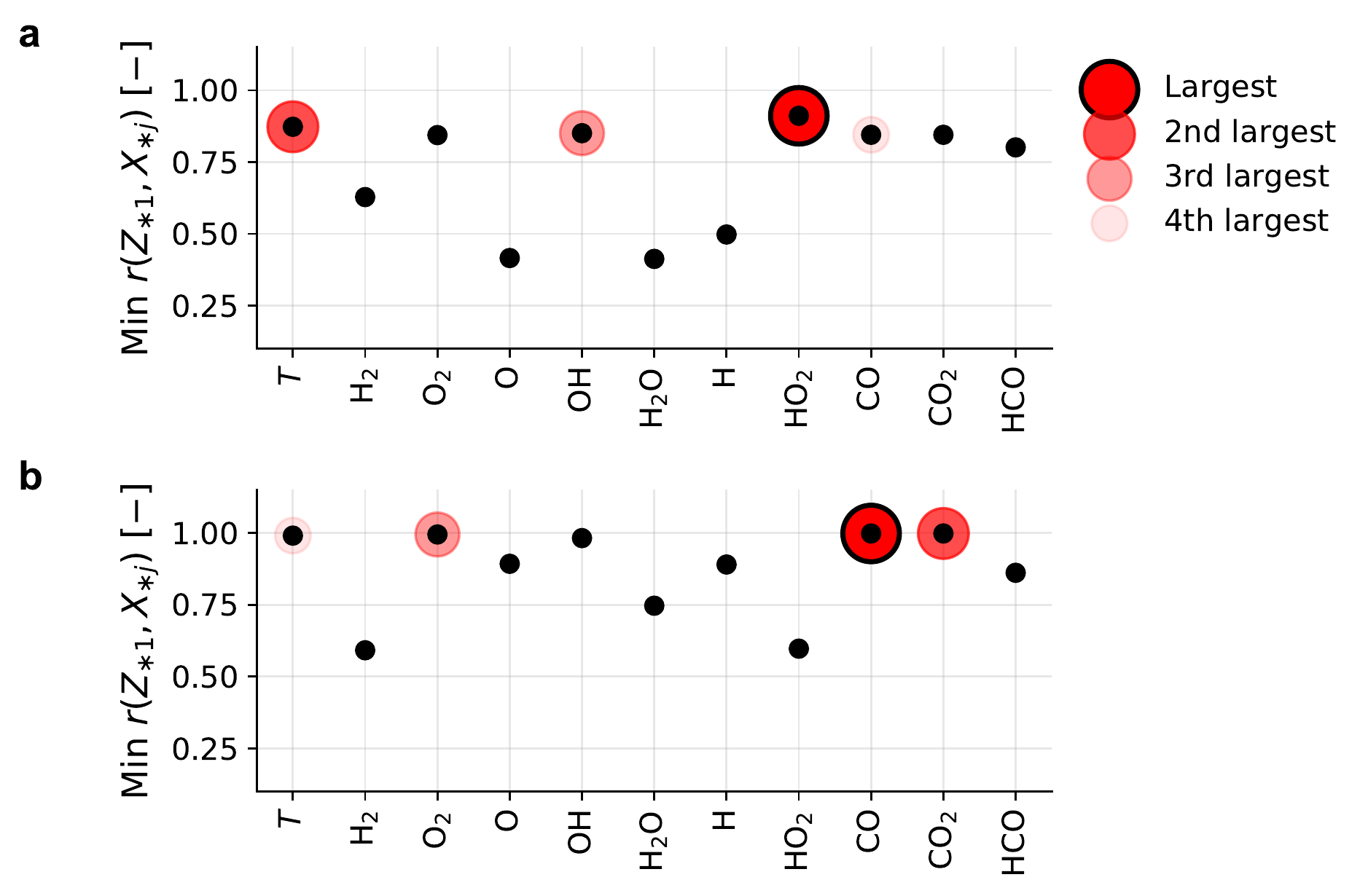}
\caption{The minimum $r$ values between the first local PC and each local state variable, $X_{*j}$, when clustering the HR dataset for \textbf{syngas/air} combustion with VQPCA with an increasing number of clusters, $k$, and using $q$=1. 
(a) Local PCA was performed using Auto scaling on a dataset $\mathbf{X} = \big[T, Y_i \big]$.
(b) Local PCA was performed using Pareto scaling on a dataset $\mathbf{X} = \big[Y_i \big]$. 
In (a) and (b), clusters were initialized by binning $Y_{H_2O}$. Legend applies to (a) and (b).}
\label{fig:HR-clustering-minimum-correlations-COH2}
\end{figure}

Similarly as we have done for the EQ dataset, we can compute analogous correlations but for the HR data clustering obtained from binning the \ch{H2O} mass fraction vector. These results are shown in Fig.~\ref{fig:HR-clustering-correlations} with dashed lines. We again observe smaller correlations in terms of $r$ and dCor as compared to clustering with the VQPCA algorithm, with the exception of VQPCA clustering with $k$=2 in Fig.~\ref{fig:EQ-clustering-correlations}a. Even though correlations associated with Pareto scaling are nearing 100\%, we see smaller correlations for all $k$ when we bin the \ch{H2O} mass fraction vector as compared to clustering using VQPCA. This suggests that the data preprocessing case based on Pareto scaling of the species mass fractions is more appropriate when optimized progress variables need to be formulated.

\begin{figure*}[t]
\centering\includegraphics[width=\textwidth]{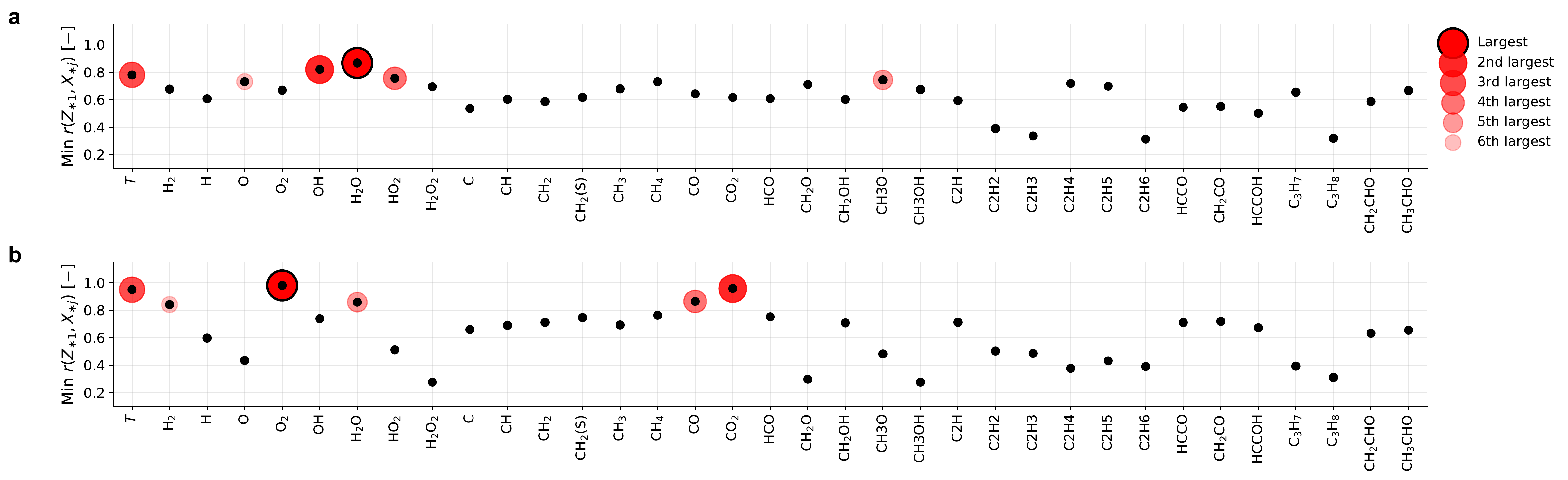}
\caption{The minimum $r$ values between the first local PC and each local state variable, $X_{*j}$, when clustering the HR dataset for \textbf{methane/air} combustion with VQPCA with an increasing number of clusters, $k$, and using $q$=1.
(a) Local PCA was performed using Auto scaling on a dataset $\mathbf{X} = \big[T, Y_i \big]$.
(b) Local PCA was performed using Pareto scaling on a dataset $\mathbf{X} = \big[Y_i \big]$.
In (a) and (b), clusters were initialized by binning $Y_{H_2O}$. Legend applies to (a) and (b).}
\label{fig:HR-clustering-minimum-correlations-CH4}
\end{figure*}

In addition, Figs.~\ref{fig:HR-clustering-minimum-correlations-H2}-\ref{fig:HR-clustering-minimum-correlations-CH4} report the minimum $r$ values for correlation between the first local PC and each local state variable in the HR datasets across three considered fuels: hydrogen, syngas and methane. The minimum $r$ was searched across $k$ changing from 2 to 4; Figs.~\ref{fig:HR-clustering-minimum-correlations-H2}-\ref{fig:HR-clustering-minimum-correlations-CH4} are thus a more compact presentation of correlations with respect to Fig.~\ref{fig:HR-clustering-correlations}. With the red shaded outlines, we mark the first few largest $r$ values. To summarize the findings from Figs.~\ref{fig:HR-clustering-minimum-correlations-H2}-\ref{fig:HR-clustering-minimum-correlations-CH4}, when Auto scaling is used on a dataset $\mathbf{X} = \big[T, Y_i \big]$, the highest correlations are observed for
\begin{itemize}
\item \ch{H2O2}, \ch{HO2}, $T$, \ch{H} for hydrogen/air,
\item \ch{HO2}, $T$, \ch{OH}, \ch{CO} for syngas/air,
\item \ch{H2O}, \ch{OH}, $T$, \ch{HO2}, \ch{CH3O}, \ch{O} for methane/air,
\end{itemize}
and when Pareto scaling is used on a dataset $\mathbf{X} = \big[Y_i \big]$, the highest correlations are observed for
\begin{itemize}
\item \ch{O2}, \ch{H2O}, $T$, \ch{H2} for hydrogen/air,
\item \ch{CO}, \ch{CO2}, \ch{O2}, $T$ for syngas/air,
\item \ch{O2}, \ch{CO2}, $T$, \ch{CO}, \ch{H2O}, \ch{H2} for methane/air.
\end{itemize}
For methane/air combustion we neglect \ch{NOx} species from this analysis. The species mass fractions exhibiting highest correlations are consistent with variables typically used in the literature to construct progress variables \citep{van2002modelling, barlow2017defining, vasavan2020novel, chung2021data, dalakoti2021priori}. This is especially the case when Pareto scaling is used on a dataset $\mathbf{X} = \big[Y_i \big]$. This result is consistent with the work done in \cite{parente2013principal}, where the authors reported that Pareto scaling emphasizes major and stable species in the definition of the first global PCs. Correlation analysis as we have performed here, can thus help in formulating optimized local $c$ as a linear combination of state variables \cite{najafiyazdi2012}.

\subsection{The high-fidelity DNS dataset: data-aided interpretation and relation with the training manifold} \label{sec:DNS}

\begin{figure*}[t ]
\centering\includegraphics[width=\textwidth]{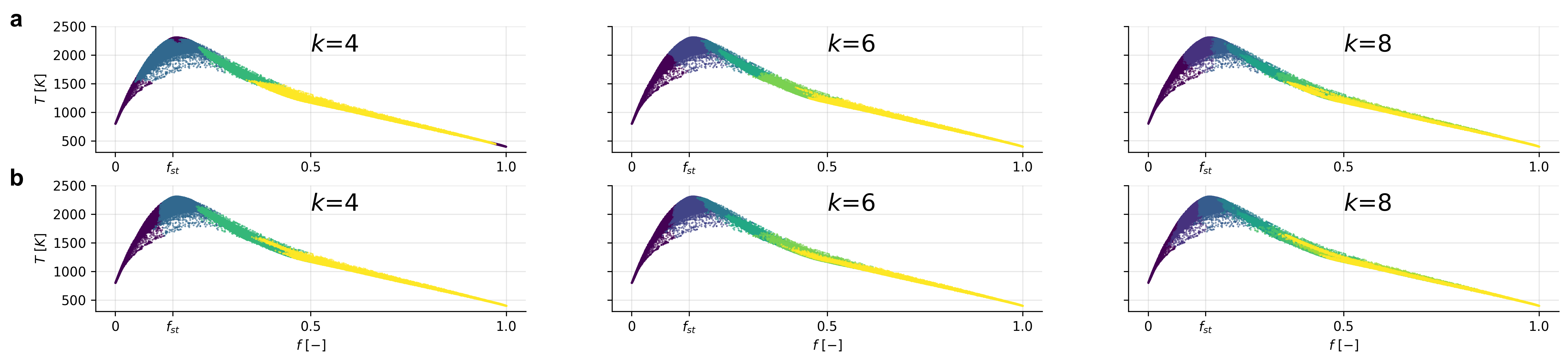}
\caption{The result of VQPCA clustering of the DNS dataset to $k$ clusters using $q$=3.
(a) Local PCA was performed using Auto scaling on a dataset $\mathbf{X} = \big[T, Y_i \big]$.
(b) Local PCA was performed using Pareto scaling on a dataset $\mathbf{X} = \big[Y_i \big]$. 
In (a) and (b), clusters were initialized by binning $f$.}
\label{fig:DNS-clustering}
\end{figure*}

\begin{figure}[h!]
\centering\includegraphics[width=7.5cm]{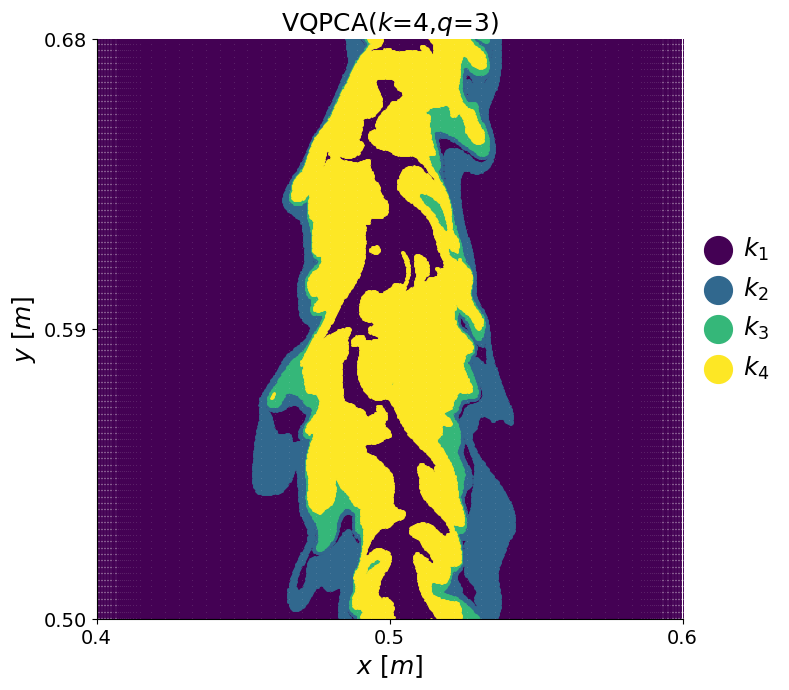}
\caption{The result of clustering the DNS dataset using VQPCA($k$=4,$q$=3) plotted in the physical space, $x$-$y$. Local PCA was performed using Auto scaling on a dataset $\mathbf{X} = \big[T, Y_i \big]$. Clusters were initialized by binning $f$.}
\label{fig:DNS-clustering-k4}
\end{figure}

Thus far, we have used our physical knowledge of the system to guide the selection of $q$. Since the intrinsic parameterization of the DNS dataset is not known \textit{a priori}, we first search the space of possible pairs ($k$,$q$) to apply on the DNS dataset. Our approach for obtaining a reasonable pair ($k$,$q$) is based on measuring the reconstruction quality from local PCA. We look at how closely the conditional means of five selected reconstructed state variables ($T$, $Y_{H_2O}$, $Y_{O_2}$, $Y_{CO_2}$ and $Y_{n-C_7H_{16}}$) approximate their true conditional means. The choice for $k$ and $q$ is made once the selected variables conditioned on $f$ are reconstructed reasonably well. We denote the true conditional mean of variable $X_{*j}$ as $\langle X_{*j} | f \rangle_t$ and the reconstructed conditional mean as $\langle X_{*j} | f \rangle_r$. To define the conditioning variable, we generate 20 bins in the space of $f$. Bins have varying sizes such that they are tighter close to $f_{st}$ and larger close to $f=0$ and $f=1$. For the $j^{\text{th}}$ selected variable, $X_{*j}$, we measure the $j^{\text{th}}$ coefficient of determination for the conditional mean as
\begin{equation}
R^2_{j} = 1 - \frac{\sum_{i=1}^{20} \Big( \langle X_{*j} | f \rangle_{t, i} - \langle X_{*j} | f \rangle_{r, i} \Big)^2}{\sum_{i=1}^{20} \Big( \langle X_{*j} | f \rangle_{t, i} - \text{mean}(\langle X_{*j} | f \rangle_{t, i}) \Big)^2}.
\end{equation}
We then find the minimum $R_j^2$ that happened for any of the five selected variables, which becomes our final reconstruction quality measure, $R^2$. We set the threshold constraint for selecting the smallest $q$ when $R^2 \geq 0.99$. If locally linear regions of the low-dimensional manifold are found, a multi-linear model with a correct dimensionality would reconstruct the data perfectly (recall Fig.~S1). With the physical scatter present in the DNS data, we do not expect $R^2$ to be equal to unity with $q < Q$. Supplementary Fig.~S16 shows a colormap of the $R^2$ values over a Cartesian product $k \times q$, where $k = [2, 3, \dots, 8]$ and $q = [1, 2, \dots, 10]$. The highest $R^2$ values happen for $q \geq 3$, suggesting that the DNS dataset should be approximated with at least three-dimensional manifolds. The black cross in Fig.~S16 marks the maximum $R^2$ which happened for ($k$=7,$q$=10). The black circle marks the smallest dimensionality $q$ for which $R^2$ is at least 0.99 which happened for ($k$=5,$q$=3). In addition, Fig.~S17 shows a comparison of conditional means for the case marked with the black circle (Fig.~S17a) and with the black cross (Fig.~S17b). We observe that both cases reconstruct the conditional means reasonably well. We thus select $q$=3 for subsequent DNS data analysis.

Fig.~\ref{fig:DNS-clustering} shows the result of VQPCA clustering of the DNS dataset to $k$=4, $k$=6 and $k$=8 clusters using $q$=3. The clustering solutions are plotted in the $f$-$T$ space. Similarly as before, we compare two preprocessing scenarios. In Fig.~\ref{fig:DNS-clustering}a, local PCA was performed using Auto scaling on a dataset $\mathbf{X} = \big[T, Y_i \big]$. In Fig.~\ref{fig:DNS-clustering}b, local PCA was performed using Pareto scaling on a dataset $\mathbf{X} = \big[Y_i \big]$. In both cases, clusters were initialized by binning the $f$ vector. In Fig.~\ref{fig:DNS-clustering-k4}, we further visualize one particular clustering solution, VQPCA($k$=4,$q$=3), corresponding to Auto scaling with $\mathbf{X} = \big[T, Y_i \big]$ in the physical space, $x$-$y$. We take this particular solution for subsequent detailed analysis. 

We begin with the qualitative interpretation of clusters seen in Fig.~\ref{fig:DNS-clustering-k4}. The first cluster, $k_1$, contains the non-reacting observations from the $n$-heptane core jet and the air co-flow. To quantitatively support this evidence, we carry out Procrustes analysis between the shape of this cluster and the shapes of the spatial profiles of various chemical species. Procrustes analysis is a statistical tool to compare two geometrical shapes, $S_1$ and $S_2$, after accomplishing operations of translation, rotation, reflection, and scaling. It minimizes the sum of squared differences between the two considered shapes as an objective function \cite{gower1975generalized, kendall1989survey}. Procrustes analysis returns a dissimilarity coefficient, $d(S_1, S_2) \in [0,1]$, which is equal to zero when two shapes are identical. When the shape of $k_1$ is geometrically compared to the spatial profiles of chemical species, the minimum value obtained for $d$ is associated with the \ch{O2} mass fraction, $d(k_1,Y_{O_{2}}) = 0.34$. In contrast, the median value of $d$ obtained when the shape of $k_1$ is compared with any of the remaining species in the dataset is $0.75$. We also look at the isolated region where the fuel jet is injected (at $x \approx 0.5$m). The dissimilarity coefficient between the central section of $k_1$ and the region where the $n$-heptane mass fraction is greater than 0.3 is $d(k_1,Y_{n-C_7H_{16}}) = 0.33$.  From the analogous geometrical analysis, the second cluster, $k_2$, appears to represent the layer where oxygen starts to react and hydroxyl and oxygen radicals are formed, since $d(k_2,Y_{OH}) = 0.4$ and $d(k_2,Y_{O}) = 0.46$. In the supplementary material, we provide additional figures for a qualitative visual comparison of the shapes of clusters $k_1$ and $k_2$ and the relevant species profiles (see Figs.~S18-S19).

\begin{figure}[t]
\centering\includegraphics[width=8cm]{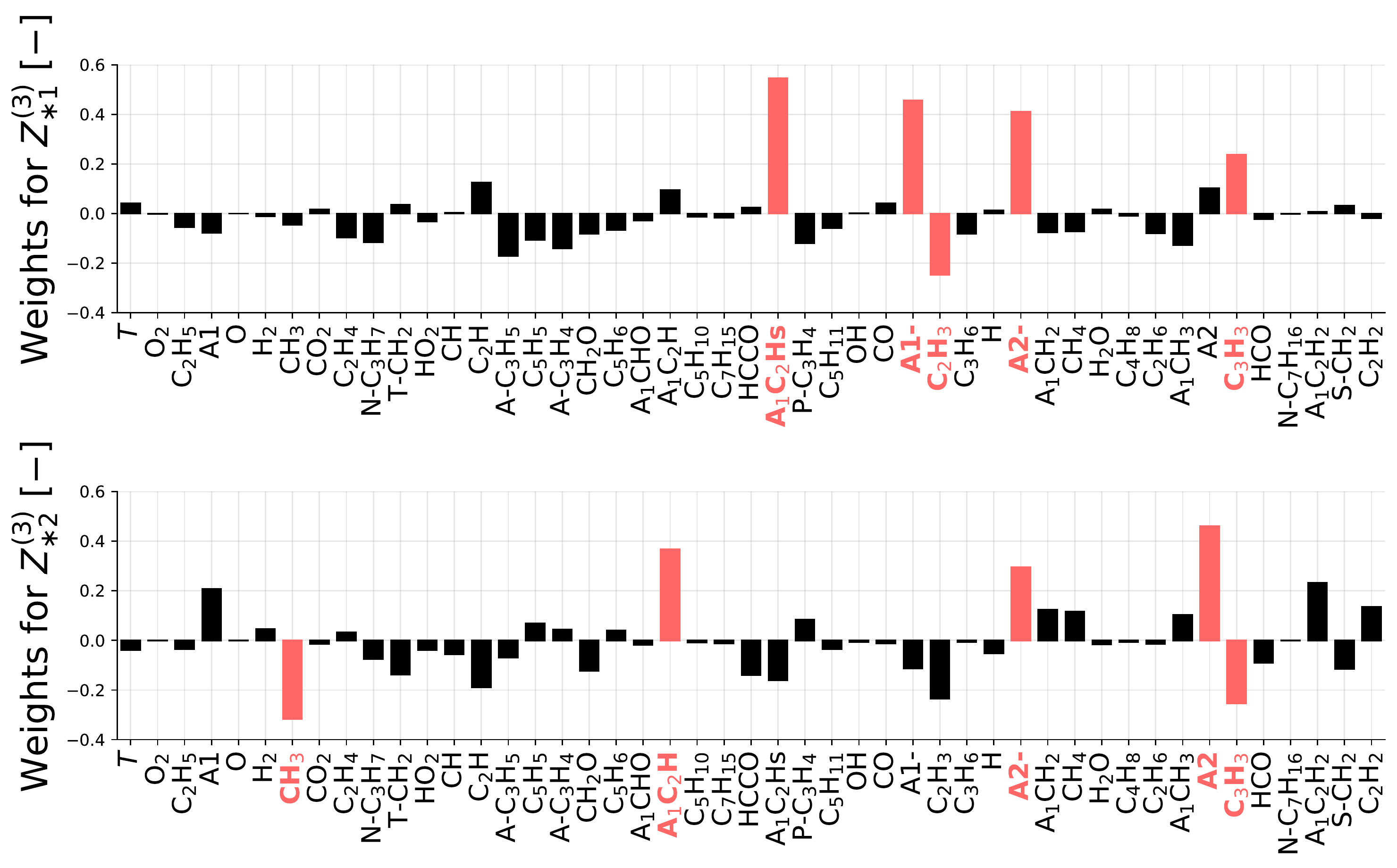}
\caption{Quantitative physical interpretation of the first two local eigenvectors from the DNS dataset in cluster $k_3$ resulting from the VQPCA($k$=4,$q$=3) solution. Local PCA was performed using Auto scaling on a dataset $\mathbf{X} = \big[T, Y_i \big]$. Clusters were initialized by binning $f$.}
\label{fig:DNS-PCs-VQPCA-k4-q3-cluster-3}
\end{figure}

The third cluster, $k_3$, is particularly interesting from the physical and chemical point of view. In Fig.~\ref{fig:DNS-PCs-VQPCA-k4-q3-cluster-3}, we examine the eigenvector weights associated with the first two local PCs in $k_3$ and we highlight the five highest weights (in terms of an absolute value) in red. Variables with the highest weights are the ones associated with the polycyclic aromatic hydrocarbons (PAH) growth and consumption. In particular, the high weight for propargyl (\ch{C3H3}) can be, from a chemical perspective, considered to be particularly coherent with the other PAHs despite not having an aromatic structure. This species is in fact responsible for the formation of aromatic compounds according to the reactions 2\ch{C3H3} = A1 and 2\ch{C3H3} = A1$-$ + \ch{H}. In addition, propargyl is also involved in the PAH growth as it is responsible for naphthalene formation according to \ch{A1CH2} + \ch{C3H3} = 2\ch{H} + A2 \cite{bisetti2012formation}. High weight for naphthalene (A2) shows up in the second PC. This can be considered an interesting finding, since the evolution of soot has been shown to be dictated by the evolution of A2 \cite{attili2016effects}. It has also been shown that the soot precursor A2 can be used as an indicator of processes that act to increase the local strain rate \cite{attili2014formation}. Cluster $k_3$ thus appears to be associated with soot formation processes, as the main PAHs and radicals involved in this process exhibit dominant weights in the first two PCs. In the supplementary Fig.~S20, we support these conclusions through a visual comparison between the shape of cluster $k_3$ and the profiles of the species that were found to be important in defining the local manifold. We compare the shape of $k_3$ in Fig.~S20a with the profiles of A1$-$, A2$-$, \ch{C3H3}, A1 and A2 in Figs.~S20b-f. We observe that the shape of $k_3$ encapsulates particularly well the regions in the flame exhibiting high mass fractions of these species.

\begin{figure}[t]
\centering\includegraphics[width=7.5cm]{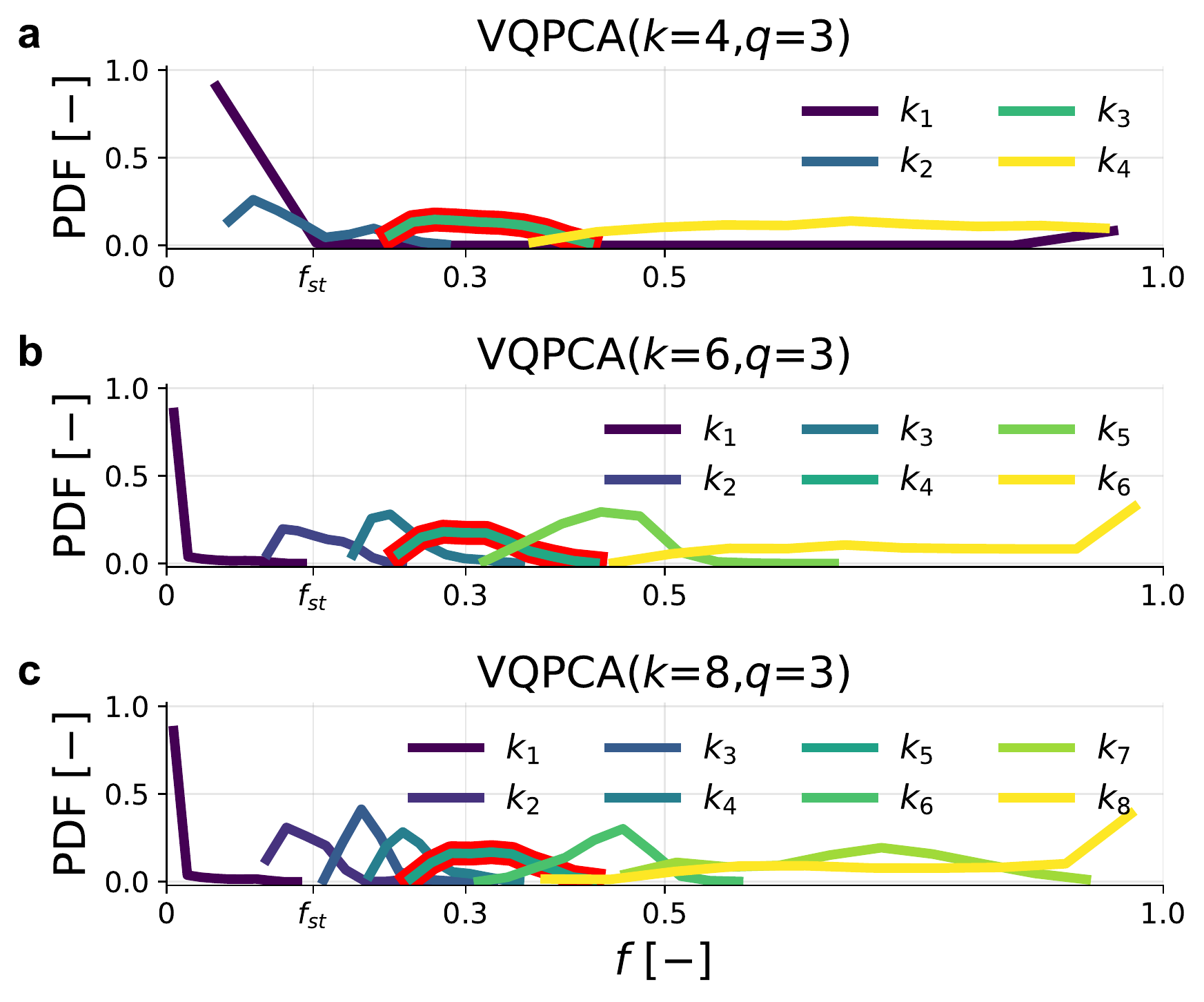}
\caption{Quantitative physical interpretation of the VQPCA clusters in the DNS dataset. We show the probability density functions (PDFs) in the $f$-space among clusters obtained using VQPCA with $q$=3 and (a) $k$=4, (b) $k$=6 and (c) $k$=8. With the red outline, we mark the cluster which captures the PAH and sooting processes in each clustering solution.
Local PCA was performed using Auto scaling on a dataset $\mathbf{X} = \big[T, Y_i \big]$. Clusters were initialized by binning $f$.}
\label{fig:DNS-PDFs}
\end{figure}

\begin{figure}[t]
\centering\includegraphics[width=8cm]{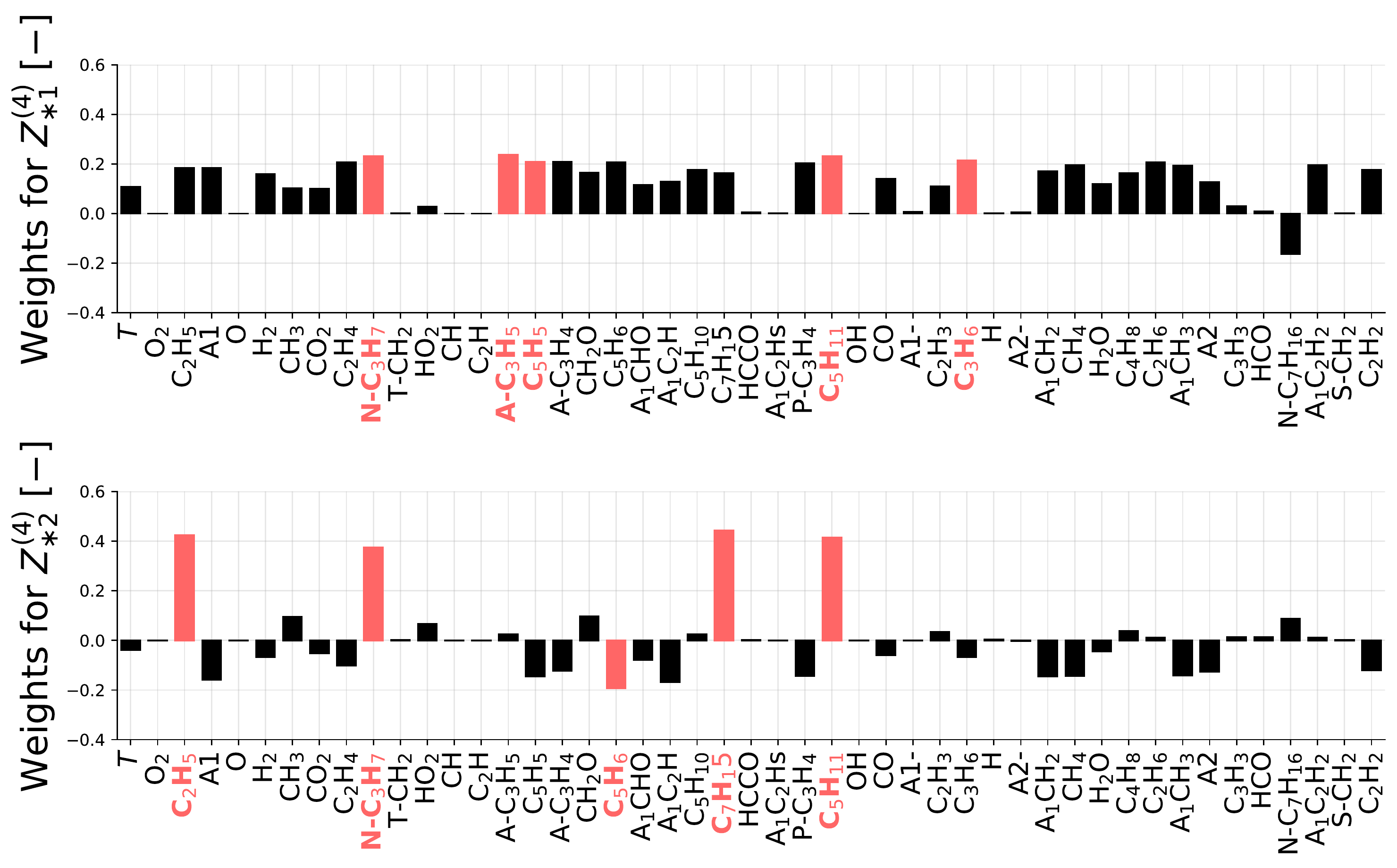}
\caption{Quantitative physical interpretation of the first two local eigenvectors from the DNS dataset in cluster $k_4$ resulting from the VQPCA($k$=4,$q$=3) solution. Local PCA was performed using Auto scaling on a dataset $\mathbf{X} = \big[T, Y_i \big]$. Clusters were initialized by binning $f$.}
\label{fig:DNS-PCs-VQPCA-k4-q3-cluster-4}
\end{figure}

For the very same chemical system and flame configuration, Attili et al. \cite{attili2014formation} observed that the peak in acetylene (\ch{C2H2}) and naphthalene mass fractions occurred at $f \approx 0.3$. Concurrently, soot was absent at smaller values of $f$ ($f < 0.3$) due to oxidation. For the DNS simulation time that we study here ($t = 15$ms), soot was also present in the entire rich side of the flame. We carry out an additional quantitative physical interpretation of cluster $k_3$ from the VQPCA($k$=4,$q$=3) solution in the $f$-space. In Fig.~\ref{fig:DNS-PDFs}a, we examine probability density functions (PDFs) in $f$ among clusters obtained with VQPCA($k$=4,$q$=3). The interval of $f$ where the flame is expected to exhibit sooting behavior is contained within $k_3$ (marked with the red outline). In addition, VQPCA seems to be always finding a cluster dedicated to PAHs and soot, even when the algorithm is applied with a different number of clusters. This is shown in Figs.~\ref{fig:DNS-PDFs}b-c, where the analogous PDFs for $f$ are reported for the VQPCA solutions using $k$=6 and $k$=8, respectively. In each case, we mark the cluster which captures the sooting processes occurring at $f \approx 0.3$ with the red outline. The supplementary Figs.~S22-S23 demonstrate the eigenvector weights in the outlined clusters from the VQPCA($k$=6,$q$=3) and the VQPCA($k$=8,$q$=3) solutions. We note the similar structure to the one presented in Fig.~\ref{fig:DNS-PCs-VQPCA-k4-q3-cluster-3}. This observation suggests that our local PCA approach can be generalizable for different values of $k$ with a fixed dimensionality $q$.

Finally, we look into the fourth cluster, $k_4$, which encapsulates the region close to the rich side of the flame. Following the results of \cite{attili2014formation}, we expect that this cluster can still capture the soot formation chemistry, but for conditions $f \rightarrow 1$. Fig.~\ref{fig:DNS-PCs-VQPCA-k4-q3-cluster-4} shows the eigenvector weights for the first two PCs in $k_4$. We note a qualitatively different PC structure to the one coming from $k_3$. In particular, the two PCs do not exhibit high weights for A2; this is in contrast to what we observe in Fig.~\ref{fig:DNS-PCs-VQPCA-k4-q3-cluster-3}. Since $k_4$ spans the range in $f$ from $f \approx 0.35$ to 1.0 (see Fig.~\ref{fig:DNS-PDFs}a), we hypothesize, that cluster $k_4$ represents regions where A2 does not necessarily form due to high dissipation rates. Both acetylene and naphthalene have been shown to be sensitive to scalar dissipation rate in laminar \citep{bohm2001pah} and turbulent flames \citep{bisetti2012formation, attili2014formation}. The presence of A2 within cluster $k_4$ can be mostly due to transport of soot towards rich side of the flame \cite{attili2014formation}.

We now extend the application of our approach to the SLF dataset for $n$-heptane/air combustion which is essentially a one-dimensional case corresponding to the analyzed DNS dataset. Attili et al. \cite{attili2016effects} studied the effect of differential diffusion on accurate representation of the non-premixed sooting flames. The authors made an interesting observation that unity Le number assumption should be used in flamelet models to better represent turbulent combustion. In this section, we apply our local PCA approach to two flamelet models as introduced in \S\ref{sec:data-sets}: one that assumes unity Le and one that assumes mixture-averaged diffusion. Our goal is to show that we arrive at the same observation as Attili et al. \cite{attili2016effects} but in a purely data-driven way. We use the DNS VQPCA clustering solutions to inform the SLF dataset clustering. Our procedure is to use $q$=3 and use the local eigenvectors from the DNS dataset to partition the SLF data \cite{d2020adaptive}. This gives us a chance to compare local PCA performed on the DNS and the SLF data on equal terms.

\begin{figure}[t]
\centering\includegraphics[width=7.5cm]{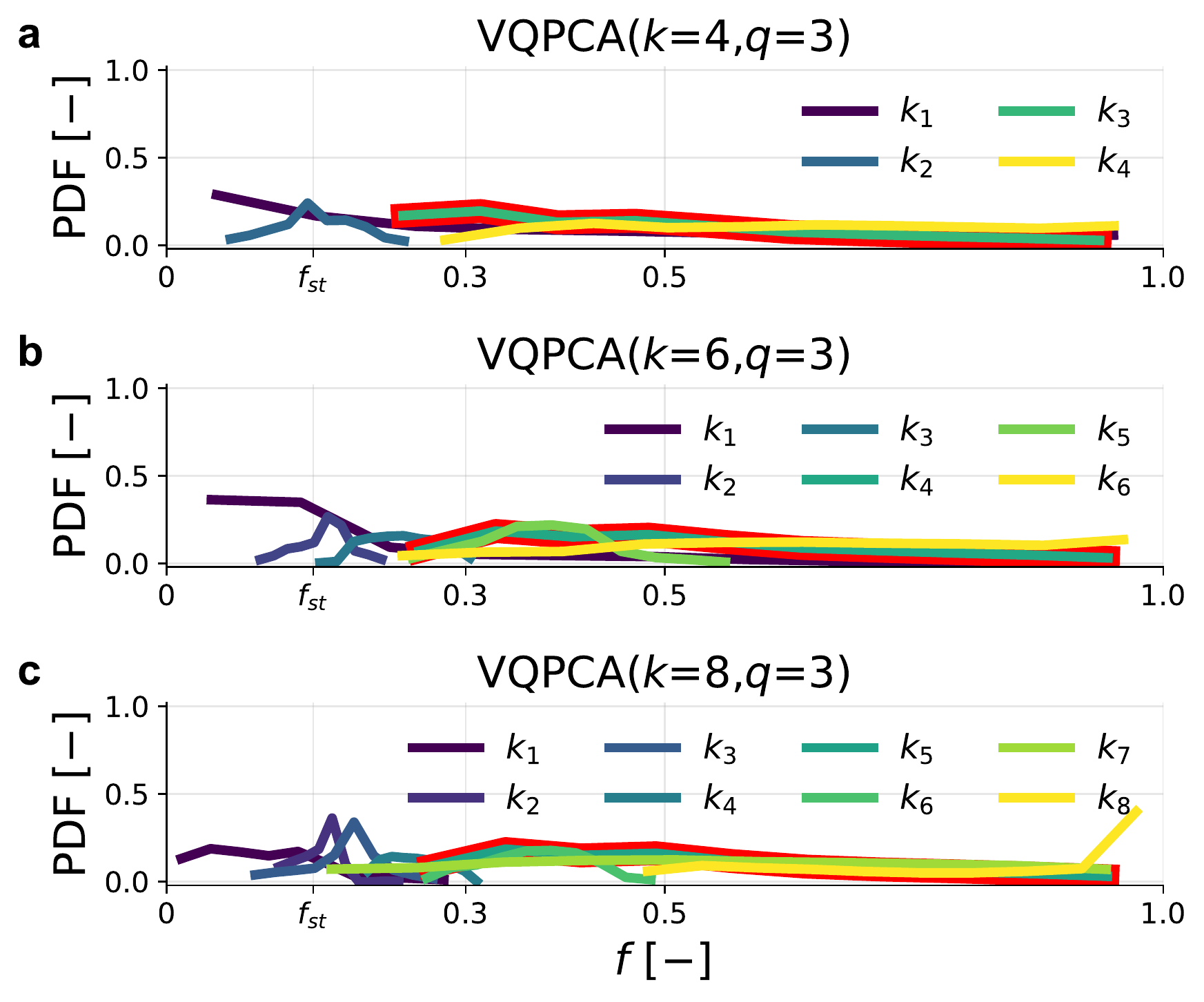}
\caption{Quantitative physical interpretation of the VQPCA clusters in the SLF dataset in the $f$-space. \textbf{We use the unity Le assumption}. We show the probability density functions (PDFs) in $f$ among clusters obtained using VQPCA with $q$=2 and (a) $k$=4, (b) $k$=6 and (c) $k$=8. With the red outline, we mark the cluster which captures the PAH and sooting processes in each clustering solution.
Local PCA was performed using Auto scaling on a dataset $\mathbf{X} = \big[T, Y_i \big]$. Clusters were initialized by binning $f$.}
\label{fig:SLF-unity-Le-PDFs}
\end{figure}

\begin{figure}[t]
\centering\includegraphics[width=8cm]{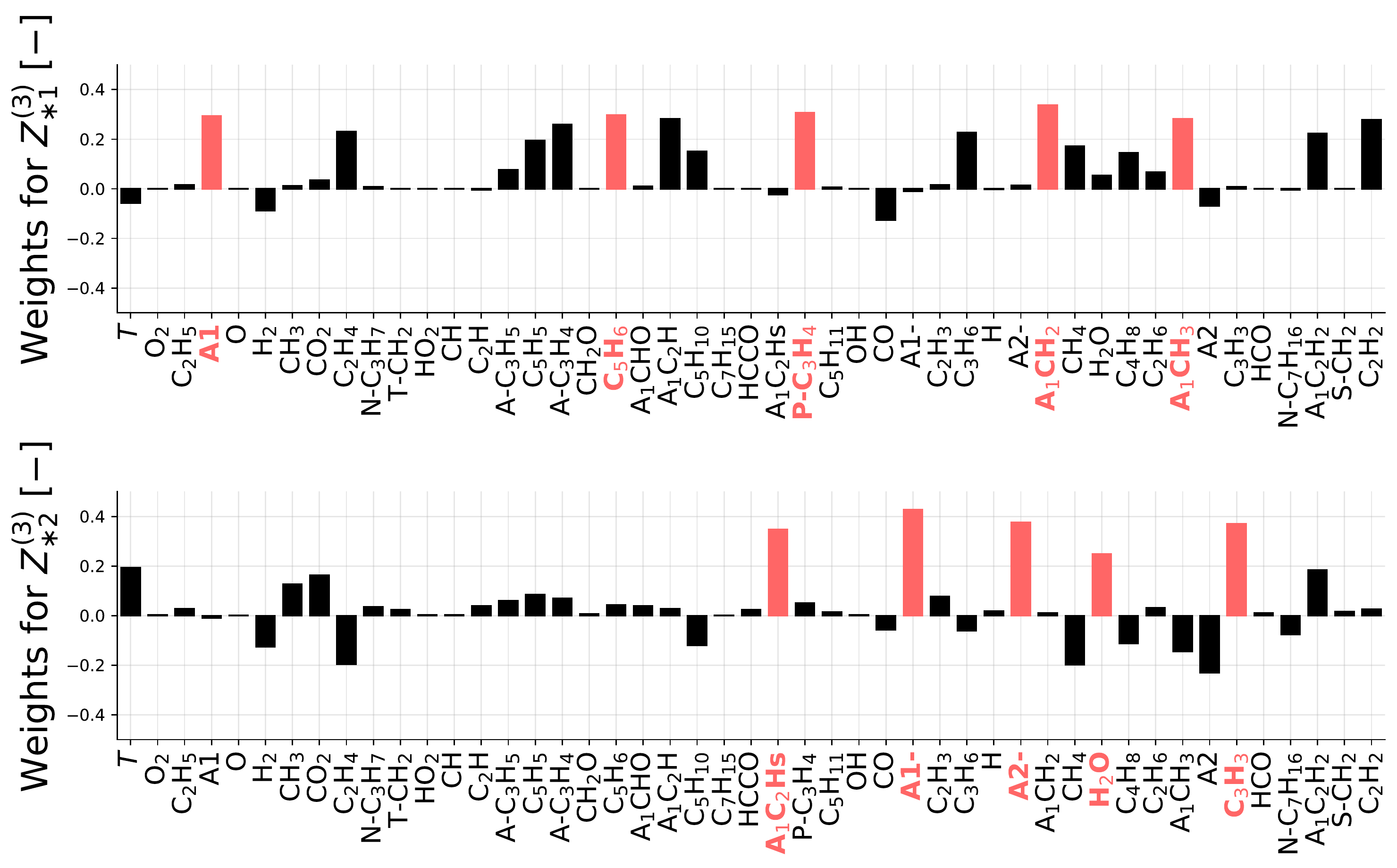}
\caption{Quantitative physical interpretation of the first two local eigenvectors from the SLF dataset in cluster $k_3$ resulting from the VQPCA($k$=4,$q$=3) solution. \textbf{We use the unity Le assumption}. Local PCA was performed using Auto scaling on a dataset $\mathbf{X} = \big[T, Y_i \big]$. Clusters were initialized by binning $f$.}
\label{fig:SLF-unity-Le-PCs-VQPCA-k4-q3-cluster-3}
\end{figure}

In an analogous way as for the DNS dataset, we compute the PDFs in the $f$-space. Fig.~\ref{fig:SLF-unity-Le-PDFs} shows the PDFs for VQPCA clustering solutions obtained for $k$=4, $k$=6 and $k$=8 clusters when the unity Le assumption is used. We note that also for this SLF dataset, we can identify a cluster associated with the soot formation processes occurring at $f \approx 0.3$. In each case, we mark that cluster with the red outline. We note that in contrast to what we observed for the DNS dataset, the soot-related cluster spans a wider range in $f$. In each case in Fig.~\ref{fig:SLF-unity-Le-PDFs}, we observe that the outlined cluster contains region from $f$ slightly lower than 0.3, all the way to $f=1.0$. We can take a closer look at the VQPCA($k$=4,$q$=3) solution, in an analogous way to what we have done for the DNS dataset. In Fig.~\ref{fig:SLF-unity-Le-PCs-VQPCA-k4-q3-cluster-3}, we examine the eigenvector weights associated with the first two local PCs in $k_3$ from the VQPCA($k$=4,$q$=3) solution. For both PCs, we mark the five highest weights (in terms of an absolute value) in red. The first PC exhibits highest weights for variables that did not show up in the PC structure for the DNS dataset. However, weights on the second PC are highest for \ch{A1C2Hs}, A1$-$,  A2$-$, \ch{H2O} and \ch{C3H3}. This PC thus resembles the first PC from $k_3$ in the DNS dataset (cf. with Fig.~\ref{fig:DNS-PCs-VQPCA-k4-q3-cluster-3}). The supplementary Figs.~S25-S26 demonstrate the eigenvector weights in the outlined clusters from the VQPCA($k$=6,$q$=3) and the VQPCA($k$=8,$q$=3) solutions. We note that in each case, we can identify a PC with high weights for A1$-$, A2$-$ and \ch{C3H3}, although this behavior switches to the second PC, in contrast to what we observed for the DNS dataset. The shared similarities between the SLF and the DNS dataset suggest that local manifold interpretations can be carried over from simpler systems to more complex ones. This fact can be used as a tool to gain initial understanding of the PCA parameterization from the training manifold, before applying the approach to larger and more complex high-fidelity datasets. Finally, we note that for the SLF dataset we have less observations corresponding to $f \approx 0$ than we had for the DNS dataset (cf. regions $f \approx 0$ in Fig.~\ref{fig:DNS-PDFs} and Fig.~\ref{fig:SLF-unity-Le-PDFs}). Interestingly, this did not prohibit finding parameterization tied to soot formation processes, which might have been the case if global PCA was applied instead, due to much oversampled region close to $f \approx 0$ in the DNS data. 

\begin{figure}[t]
\centering\includegraphics[width=7.5cm]{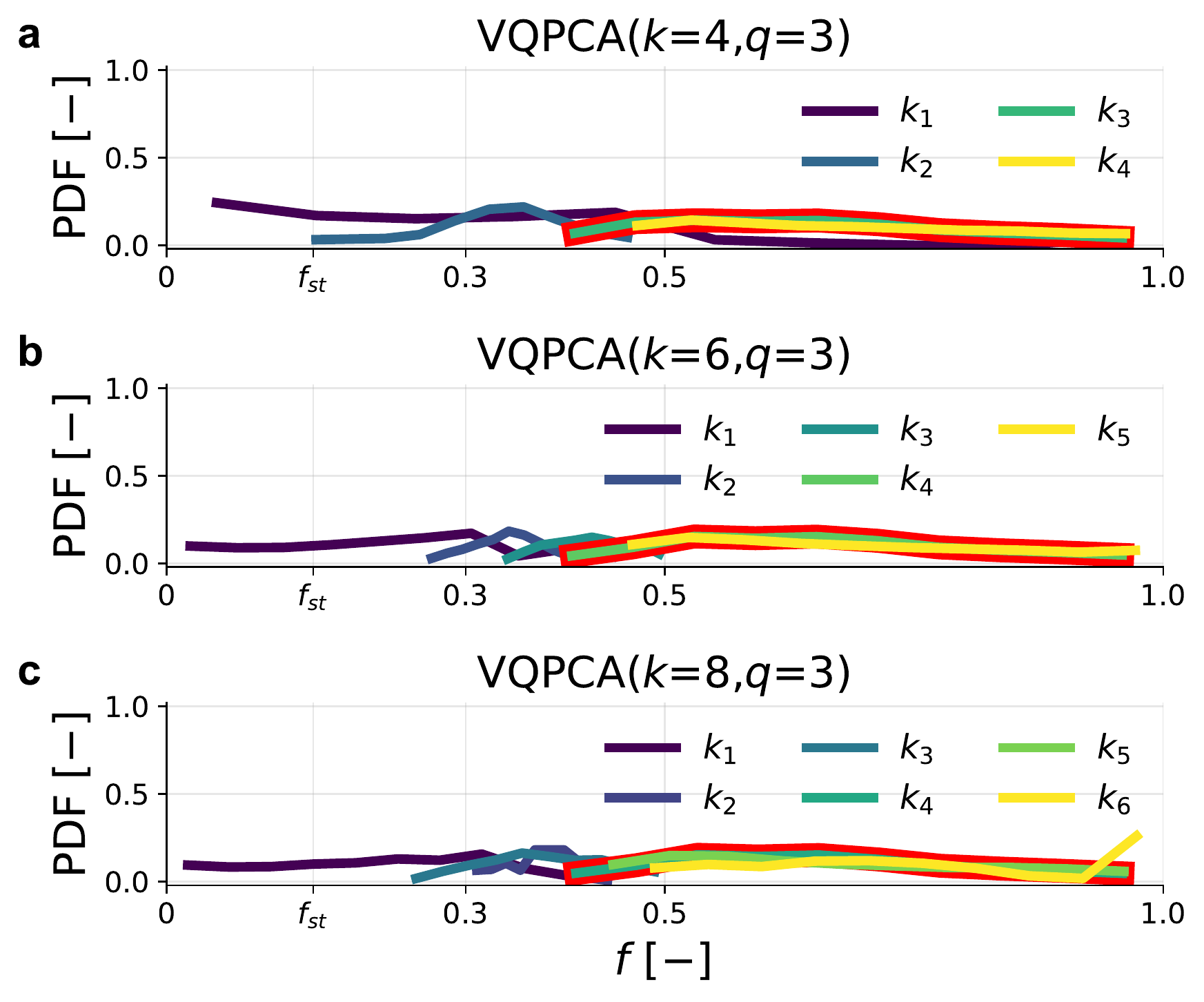}
\caption{Quantitative physical interpretation of the VQPCA clusters in the SLF dataset in the $f$-space. \textbf{We use the non-unity Le assumption}. We show the probability density functions (PDFs) in $f$ among clusters obtained using VQPCA with $q$=2 and (a) $k$=4, (b) $k$=6 and (c) $k$=8. With the red outline, we mark the cluster which captures the PAH and sooting processes in each clustering solution.
Local PCA was performed using Auto scaling on a dataset $\mathbf{X} = \big[T, Y_i \big]$. Clusters were initialized by binning $f$.}
\label{fig:SLF-non-unity-Le-PDFs}
\end{figure}

\begin{figure}[t]
\centering\includegraphics[width=8cm]{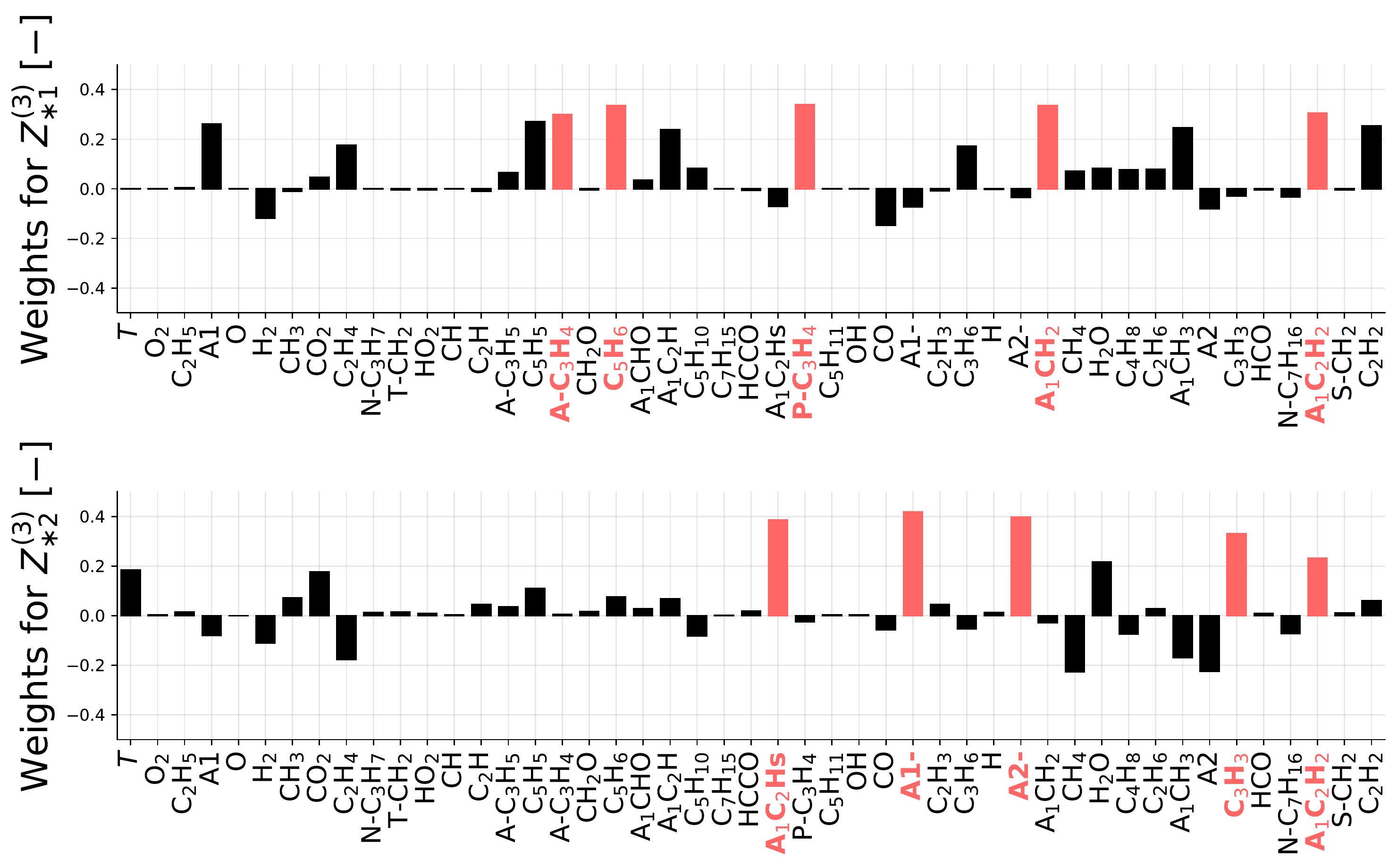}
\caption{Quantitative physical interpretation of the first two local eigenvectors from the SLF dataset in cluster $k_3$ resulting from the VQPCA($k$=4,$q$=3) solution. \textbf{We use the non-unity Le assumption}. Local PCA was performed using Auto scaling on a dataset $\mathbf{X} = \big[T, Y_i \big]$. Clusters were initialized by binning $f$.}
\label{fig:SLF-non-unity-Le-PCs-VQPCA-k4-q3-cluster-3}
\end{figure}

Fig.~\ref{fig:SLF-non-unity-Le-PDFs} shows the analogous PDFs but for the SLF with the mixture-averaged differential diffusion model. We note different probability distributions when non-unity Le assumption is used as compared to the flamelet with unity Le assumption. Primarily, clustering this SLF dataset based on local eigenvectors coming from the DNS data caused some too small clusters to disappear in the VQPCA($k$=6,$q$=3) and VQPCA($k$=8,$q$=3) cases. Thus, in Fig.~\ref{fig:SLF-non-unity-Le-PDFs}b the actual number of clusters found is five, and in Fig.~\ref{fig:SLF-non-unity-Le-PDFs}c it is six. This behavior suggests that the clustering solution obtained from the DNS data is less adequate to partition the SLF dataset with the non-unity Le assumption. The cluster associated with soot formation for each VQPCA clustering solution is highlighted in red. We observe that in each case, the soot cluster does not encapsulate the region $f \approx 0.3$. The main difference between the unity and non-unity Le assumption from the perspective of the soot cluster is that in the unity Le, this cluster spans the range from $f \approx 0.3$ to $f$=1.0. In the non-unity Le assumption, the soot cluster spans the range from $f \approx 0.4$ to $f$=1.0. This suggests, that with the more accurate differential diffusion model, PAHs show up in the PC structure due to their transport towards the rich side of the flame. Conversely, with the unity Le assumption, PAHs might have showed up in the PC structure both due to their formation at $f \approx 0.3$, and due to transport. In Fig.~\ref{fig:SLF-non-unity-Le-PCs-VQPCA-k4-q3-cluster-3}, we present the eigenvector weights associated with the first two local PCs in $k_3$ from the current VQPCA($k$=4,$q$=3) solution. In the second PC, we can still identify parameterization based on the PAH species, with the highest weights for \ch{A1C2Hs}, A1$-$,  A2$-$, \ch{C3H3} and \ch{A1C2H2}. The supplementary Figs.~S25-S26 demonstrate the eigenvector weights in the outline clusters from the VQPCA($k$=6,$q$=3) and the VQPCA($k$=8,$q$=3) solutions. We note that in each case, we can identify a PC with high weights for \ch{A1C2Hs}, A1$-$,  A2$-$, \ch{C3H3} and \ch{A1C2H2}, and similarly as in the unity Le SLF, this behavior can be seen in the second PC.

\section{Conclusions}
 
Dimensionality reduction techniques can be used to obtain effective parameterizations of complex systems with many degrees of freedom. Among many available linear and nonlinear reduction techniques, PCA is widely used in various research fields. PCA allows to compute parameterizations directly from the training data. Areas of research strongly populated with large amounts of data can particularly benefit from this data-driven approach. In this work, we present an application of local PCA to simplify the description of high-dimensional reacting flows datasets. We perform a bottom-up exercise in interpreting local PCs. We demonstrate that applying PCA in local clusters (a locally linear approach) we can retrieve the inherent parameters used to generate data from simple combustion models: BS, EQ and HR. The local parameters can be thought of as a more compact representation of the original thermo-chemical variables. For a much more complex DNS dataset, we show that under certain settings, physical meaning can be attributed to the obtained parameters. We analyze a turbulent non-premixed sooting flame DNS and show the potential to link the local parameters with soot formation processes. Finally, we demonstrate that local PCA parameterizations can be shared between the DNS data and a simpler SLF model case corresponding to the systems simulated in the DNS. We perform a data-driven demonstration that our local PCA approach can suggest which differential diffusion model is more adequate to represent a high-fidelity non-premixed turbulent flame simulation. The results reported in this work suggest that when locally linear manifolds are anticipated in the data, clustering based on linear reconstruction error minimization can be a good choice.

\section*{Acknowledgements}

The research of the first author is supported by the F.R.S.-FNRS Aspirant Research Fellow grant.
GDA has received funding from the Fonds National de la Recherche Scientique (F.R.S.-FNRS) through a FRIA fellowship.
Aspects of this material are based upon work supported by the National Science Foundation under Grant No.~1953350.
This project has received funding from the European Research Council (ERC) under the European Union’s Horizon 2020 research and innovation program under grant agreement no.~714605.

\section*{Conflict of interest}

There are no known conflicts of interest associated with this publication.

\section*{Author contributions}

Conceptualization: K.Z.; G.D.; A.A.; A.C.; J.C.S.; A.P. Data curation: K.Z.; G.D.; A.A. Formal analysis: K.Z.; G.D.;  Methodology: A.A.; A.C.; J.C.S.; A.P. Supervision: A.A.; A.C.; J.C.S.; A.P. Visualization: K.Z.; G.D. Writing - original draft: K.Z.; G.D. Writing - review \& editing: K.Z.; G.D.; A.A.; A.C.; J.C.S.; A.P. All authors approved the final submitted draft.

\section*{Available code}

All code used to produce the results can be found in the form of Jupyter notebooks in the public GitHub repository: \\
\href{https://github.com/kamilazdybal/local-manifold-learning}{\texttt{https://github.com/kamilazdybal/local-manifold-learning}} \\
The training datasets were generated using \textbf{Spitfire} software \cite{Spitfire}, available at:\\
\href{https://github.com/sandialabs/Spitfire}{\texttt{https://github.com/sandialabs/Spitfire}}\\
and \textbf{Cantera} software \cite{Cantera}.
Local PCA results were produced using \textbf{PCAfold} software \cite{PCAfold}, more information can be found at:\\
\href{https://pcafold.readthedocs.io}{\texttt{https://pcafold.readthedocs.io}}.\\
VQPCA implementation available at:\\
\href{https://github.com/burn-research/reduced-order-modelling}{\texttt{https://github.com/burn-research/reduced-order-modelling}}\\
was used. \\
Distance correlation was computed using \textbf{dcor} Python package available at:\\ 
\href{https://github.com/vnmabus/dcor}{\texttt{https://github.com/vnmabus/dcor}}.

\end{document}